\documentclass[11pt]{article}
\usepackage{graphicx}
\usepackage{color}
\usepackage{amsmath,amssymb,amsfonts,latexsym}
\usepackage{makecell}
\begin{document}

\begin{center}
\textbf{\Large Maximum Likelihood and Bayesian Estimation \\[2mm]
for State-Space Models Using the Non-Gaussian Filter}

\vspace{8mm}
{\large Genshiro Kitagawa}\\[2mm]
Tokyo University of Marine Science and Technology\\[-1mm]
and\\[-1mm]
The Institute of Statistical Mathematics

\vspace{3mm}
{\today}
\end{center}

\noindent
\textbf{Abstract:}
The non-Gaussian filter provides a deterministic numerical method 
for nonlinear and non-Gaussian state-space models, 
but its application has long been limited due to the
computational cost of numerical integration. 
Advances in computing power and memory capacity have substantially reduced 
this limitation for low and moderate dimensional models. This paper
re-examines the non-Gaussian filter and demonstrates its usefulness for
maximum likelihood estimation and Bayesian inference. Numerical experiments with
linear, nonlinear and radar-tracking models show that log-likelihood 
obtained by non-Gaussian filter is smooth and can be optimized reliably, whereas
the ones obtained by particle filter are affected strongly by Monte Carlo
variability even with many particles. Bayesian estimation is performed
using a self-organizing state-space model, in which unknown parameters
are incorporated into the state vector and estimated jointly with the
latent states. These results demonstrate that the current computing technology has
renewed the practical value of deterministic filtering for statistical
inference in state-space models.

\vspace{2mm}

\section{Introduction}

State-space models provide a unified framework for modeling, estimation 
and prediction of time series with dynamic structures.
For linear Gaussian state-space models, the Kalman filter recursively
computes the predictive, filtering and smoothing distributions
(Anderson and Moore, 1979).
Consequently, many important statistical problems, including
parameter estimation, prediction, missing-data interpolation,
signal extraction, and component decomposition,
can be formulated as state estimation problems.

The Kalman filter, however, is applicable only under the assumptions of
linearity of the state-space model and Gaussianity of the system and
observation noises.
Many practical applications, involve nonlinear state-space
models and/or non-Gaussian system or observation noises, such as heavy-tailed, 
skewed and multimodal distributions.
For such problems, particle filters have become the standard approach
because of their flexibility and broad applicability
(Gordon et al., 1993; Kitagawa, 1996; Doucet et.al., 2001; Doucet and Johansen, 2011).

Despite these advantages, particle filters have an important limitation
when likelihood-based inference is required.
Since the likelihood is estimated by Monte Carlo computation, the
resulting log-likelihood surface is inherently affected by Monte Carlo noise,
making numerical optimization for maximum likelihood estimation
considerably more difficult.
Although repeated runs can reduce the Monte Carlo variance, the
convergence is slow, and the likelihood surface remains irregular even
when the Monte Carlo variance is substantially reduced.
Consequently, conventional optimization methods such as quasi-Newton
algorithms often perform poorly.

Particle filters also encounter an important difficulty in Bayesian estimation of
unknown parameters using self-organizing state-space models (Kitagawa, 1998).
In this approach, unknown parameters are incorporated into an augmented
state vector and estimated simultaneously with the system state.
Because particle filters require evolution noise for the parameter
components, they implicitly assume that the parameters evolve over time.
When the parameters are truly time-invariant, this assumption is
violated, leading to severe particle degeneracy and poor estimation
performance.

An alternative deterministic approach is the non-Gaussian filter,
originally proposed by Kitagawa (1987).
Instead of representing the posterior distribution by random particles,
the non-Gaussian filter directly approximates the probability density
functions by numerical integration on discretized state-spaces.
When it was originally proposed, this approach was considered
computationally prohibitive except for very low-dimensional problems.
However, advances in processor speed and memory capacity over the past
four decades--each by approximately nine orders of magnitude--have
dramatically changed this situation.
For state-space models with dimensions up to about four, the
computational cost has become entirely practical even on ordinary desktop
computers.

For parameter estimation, the non-Gaussian filter has
two important advantages over particle filters.
The first advantage is that the likelihood is deterministic rather than stochastic.
As a result, the resulting log-likelihood function is smooth, allowing
standard optimization algorithms to obtain accurate maximum likelihood
estimates without being affected by Monte Carlo noise.

The second advantage, the non-Gaussian filter naturally accommodates both
time-invariant and time-varying parameters in self-organizing state-space models.
Unlike particle filters, no artificial system noise is required for
constant parameters.
Moreover, when the parameters are time-invariant, the posterior
distribution of the parameters is identical to the filtering
distribution at the final observation.
Therefore, smoothing of the parameter components is unnecessary, leading
to substantial reductions in both computational cost and memory
requirements.

This paper revisits the non-Gaussian filter from a modern
computational viewpoint and demonstrates that it provides an attractive
alternative to particle filters for likelihood-based inference and
Bayesian estimation.

The main contributions of this paper are as follows.
\begin{itemize}
  \setlength{\itemsep}{0pt}
  \setlength{\topsep}{0pt}
\item A modern reassessment of the non-Gaussian filter in the context of modern
      computational environments;
\item A comparison of the non-Gaussian filter with the particle filter in terms
      of log-likelihood computation;
\item A demonstration of accurate maximum likelihood estimation using the
      deterministic log-likelihood obtained by the non-Gaussian filter;
\item A demonstration of Bayesian estimation based on self-organizing state-space
      models for both time-invariant and time-varying parameters.
\end{itemize}

The remainder of the paper is organized as follows.
Section~2 briefly reviews the non-Gaussian filter, the computation of
the log-likelihood, and the self-organizing state-space model for
simultaneous estimation of states and parameters.
Section~3 presents numerical examples of log-likelihood computation and
maximum likelihood estimation.
Section~4 investigates Bayesian estimation of states and parameters.
Finally, Section~5 summarizes the main findings.
For the convenience of the reader, the Appendix provides a brief review of the Kalman
filter and the particle filter together with their log-likelihood
computation procedures.

The objective of this paper is not to replace particle filters,
which remain indispensable for high-dimensional state-space models.
Rather, we demonstrate that advances in modern computing have made the
non-Gaussian filter a practical and competitive deterministic
alternative whenever low to moderate dimensional problems are
considered.

The paper therefore complements, rather than competes with,
particle filters by identifying situations in which deterministic
numerical integration provides significant practical advantages.

\section{State Estimation and Likelihood Computation in Nonlinear Non-Gaussian State-Space Models}

\subsection{Nonlinear Non-Gaussian State-Space Models}
Consider the following nonlinear non-Gaussian state-space model for an observed time series \(\{y_n\}\):
\begin{align}
x_n &= f(x_{n-1}) + g(v_n) \hspace{11mm} (\mbox{system model}) \label{ngauss-1-1}  \\
y_n &= h(x_n) + w_n \hspace{19mm} (\mbox{observation model}), \label{ngauss-1-2}
\end{align}
where \(x_n\) denotes the latent state with initial state $x_0 \sim p(x_0)$, $f(x)$, $g(x)$ and $h(x)$ are known, possibly nonlinear functions and the system noise $v_n$ and the observation noise $w_n$ are independent white noise sequences that follow the density functions $q(v)$ and $r(w)$, respectively. 
Unlike linear Gaussian state-space models, these distributions are not necessarily Gaussian.
The observations up to time \(j\) is denoted as $Y_{1:j} \equiv \{y_1,\ldots,y_j\}$.

For such models, the distribution of the state vector $x_n$ generally becomes non-Gaussian. Consequently, the state-space model of equations (\ref{ngauss-1-1}) and (\ref{ngauss-1-2}) is called a nonlinear non-Gaussian state-space model. 
The objective of state estimation is to determine the conditional distribution of the state vector $x_n$, given the information $Y_{1:m}$. When \(n>m\), \(n=m\) and \(n<m\),
the corresponding problems are referred to as prediction, filtering and smoothing, respectively.

The recursive formulas presented below form the basis of the
non-Gaussian filter used throughout this paper for likelihood
computation, maximum likelihood estimation and Bayesian estimation.
For linear Gaussian state-space models,
the Kalman filter recursively computes the conditional mean
and variance covariance matrix of the state vector.
For general nonlinear and non-Gaussian models, the posterior state distribution cannot be characterized solely by the first two moments and it is necessary to obtain the conditional densities for state estimation. 

For the model defined by
(\ref{ngauss-1-1}) and
(\ref{ngauss-1-2}),
the Markov property implies \( p(x_n|x_{n-1},Y_{1:n-1})=p(x_n|x_{n-1})\)
and \( p(y_n|x_n,Y_{1:n-1}) = p(y_n|x_n)\).
These properties lead directly to the recursive prediction and
filtering equations shown by Kitagawa (1987).

\medskip
\noindent
{\bf [\,Prediction\,]}
The one-step-ahead predictive distribution is given by
\begin{align}
 p(x_n |Y_{1:n-1} ) 
 = \int_{- \infty}^\infty p(x_n | x_{n-1} )
 p(x_{n-1} |Y_{1:n-1} )dx_{n-1}. \label{ngauss-1-3}
\end{align}
\noindent
{\bf [\,Filtering Update\,]}
The filtering distribution is then updated according to Bayes' theorem:
\begin{align}
 p(x_n |Y_{1:n} ) 
 = \frac{p(y_n |x_n )p(x_n |Y_{1:n-1} )}{p(y_n |Y_{1:n-1} )}, \label{ngauss-1-4}
\end{align}
where the predictive density $p(y_n|Y_{1:n-1})$ is obtained as ${\displaystyle \int p(y_n |x_n )p(x_n |Y_{1:n-1} )dx_n }$. 
The transition density \(p(x_n|x_{n-1})\) is determined by the system model (\ref{ngauss-1-1}).

An important by-product of the recursive filtering procedure is the
log-likelihood function,
\begin{align}
 \ell (\theta) = \sum_{n=1}^N \log p(y_n|Y_{1:n-1}).
\end{align}

\medskip
\noindent
{\bf [\,Smoothing\,]}
The fixed-interval smoothing distribution is obtained recursively by
\begin{align}
 p(x_n |Y_{1:N} ) 
 = p(x_n |Y_{1:n} ) \int_{- \infty}^\infty \!\frac{p(x_{n+1} |Y_{1:N} )p(x_{n+1} |x_n)}{p(x_{n+1} |Y_{1:n} )}dx_{n+1}. \label{ngauss-1-7}
\end{align}
In the right-hand side of the formula (\ref{ngauss-1-7}), $p(x_{n+1}|x_n)$ is determined by the system model (\ref{ngauss-1-1}). On the other hand, $p(x_n|Y_{1:n})$ and $p(x_{n+1}|Y_{1:n})$ are obtained by equations (\ref{ngauss-1-3}) and (\ref{ngauss-1-4}), respectively. 
Starting from the filtering distribution \(p(x_N|Y_{1:N})\),
equation (\ref{ngauss-1-7}) is applied recursively backward for \(n=N-1,\ldots,1\)
to obtain the entire fixed-interval smoothing distributions.

These recursive prediction, filtering and smoothing equations form the
foundation of the deterministic non-Gaussian filter developed in the
following subsection.

\subsection{Numerical Implementation of the Non-Gaussian Filter}

As shown in the previous subsection, the recursive equations for state estimation 
can be derived for general non-Gaussian state-space models as a natural
extension of the Kalman filter.
This unified framework makes it possible to analyze a wide variety of
time series models.
The remaining challenge is therefore how to evaluate these recursive equations 
efficiently for general nonlinear non-Gaussian models.

In this subsection, we describe the numerical implementation of the non-Gaussian filter, 
which realizes non-Gaussian filtering and smoothing by numerically approximating the
posterior probability distributions (Kitagawa 1987).
The basic idea is to approximate the non-Gaussian state densities by
simple basis functions, such as piecewise-constant (step) or piecewise-linear functions, 
and to numerically evaluate equations~(\ref{ngauss-1-3})--(\ref{ngauss-1-7}).
Although this approach was computationally prohibitive except for one-dimensional model 
when it was first proposed,
the rapid development of high-performance computers has made it
practical for low-dimensional state-space models.

In the following, the probability density functions appearing in
equations~(\ref{ngauss-1-3}), (\ref{ngauss-1-4}) and (\ref{ngauss-1-7})
are approximated by step functions (Kitagawa and Gersch, 1996; Kitagawa, 2020).
For simplicity of presentation, the state $x$ is assumed to be one dimensional.
Let the probability density function
$f(x)$ is defined on the real line, $-\infty<x<\infty$.
To construct its step function approximation,
the infinite domain is first truncated to a finite interval $[x_0,x_k]$,
which is then divided into $k$ subintervals,
$x_0<x_1<\cdots<x_k$.
The endpoints $x_0$ and $x_k$ are chosen sufficiently small and sufficiently large so that the
probability mass outside the interval is negligible.

In practice, as we will discuss in the following subsection, 
the interval is often shifted according to the location of the density.
For simplicity, however, in this subsection we assume that the grid points are fixed.
All subintervals are assumed to have equal width.
Let $ \Delta x = (x_k-x_0)/k$,
then $x_i=x_0+i\Delta x$, for $ i=0,\ldots,k$.

In the step function approximation, $f(x)$ is approximated by a constant
$f_i$ over the interval $[x_{i-1},x_i]$.
The constant height on each interval, $f_i$, is obtained by
\begin{equation}
f_i = \frac{1}{\Delta x} \int_{x_{i-1}}^{x_i} f(x)\,dx.
\end{equation}
The resulting step-function approximation is represented by the discrete representation
$ \tilde f(x) \equiv \{k; x_0,$ $\ldots,x_k; f_1,\ldots,f_k\}$.

For the numerical implementation of the non-Gaussian filtering and
smoothing algorithms, the following density functions are approximated
by step functions. These discrete representations are denoted by
$ p(x_n|Y_{1:n-1}) \approx \tilde p(x) \equiv \{k;x_0,\ldots,x_k;$ $p_1,\ldots,p_k\}$, 
$ p(x_n|Y_{1:n}) \approx \tilde f(x) \equiv \{k;x_0,\ldots,x_k;f_1,\ldots,f_k\}$, 
$ p(x_n|Y_{1:N}) \approx \tilde s(x) \equiv \{k;x_0,\ldots,x_k;s_1,\ldots,s_k\}$, 
and the system noise density
$ q(v) \approx \tilde q(v) \equiv \{2k+1;x_{-k},\ldots,x_d;$ $q_{-k},\ldots,q_k\}$.
The observation noise density $r(v)$,
on the other hand, is evaluated directly without discretization
because it appears only pointwise in the filtering equation.

Once these distributions have been discretized, the filtering and smoothing equations can
be evaluated numerically by numerical integration.
The resulting numerical procedure is referred to throughout this paper 
as the \emph{non-Gaussian filter}.

For illustration of the computational procedure, we first describe the algorithm for a
one-dimensional trend model. The extension to higher-dimensional state
space models is conceptually straightforward, although the computational
cost increases rapidly with the dimension of the state vector.

\subsubsection{Example: One-Dimensional Trend Model}
In the following, we show a procedure for numerical evaluation of the simple one-dimensional trend model; 
\begin{align}
 x_n &= x_{n-1} + v_n,\nonumber \\
 y_n &= x_n + w_n. 
\end{align}

\noindent
{\bf [\,Prediction\,]} 

The prediction step is computed as follows.
\begin{align}
 p_i \: =\: \tilde p(x_i) 
   = \int_{x_0}^{x_k} \tilde q(x_i-s) \tilde f(s) ds 
   = \sum_{j=1}^{k} \int_{x_{j-1}}^{x_j} \tilde q(x_i-s) \tilde f(s) ds 
   = \Delta x \sum_{j=1}^{k} q_{i-j} f_{j}. \label{Eq_NGF_prediction_1D}
\end{align}

\noindent
{\bf [\,Filtering\,]}

The filtering update is given by
\begin{align}
 f_i = \tilde f(x_i) = \frac{ r(y_n - x_i) \tilde p(x_i)}{C_n} 
      =\frac{ r(y_n - x_i) p_i}{C_n}, \label{Eq_NGF_filter_1D}
\end{align}
where the normalizing constant (predictive density) $C_n$ is obtained by
\begin{align}
 C_n  = \int_{x_0}^{x_k} r(y_n - x) \tilde p(x) dx 
 = \sum_{j=1}^k \int_{x_{j-1}}^{x_j} r(y_n - x) \tilde p(x) dx 
 = \Delta x\sum_{j=1}^k r(y_n - x_j) p_{j}. \label{EQ_NGF_C}
\end{align}

\noindent
\textbf{ [\,Log-likelihood of the State-Space Model\,]}

As a natural by-product of the filtering recursion, the log-likelihood is computed as
\begin{align}
   \tilde{\ell} (\theta) = \sum_{n=1}^N \log C_n. \label{Eq_NGF_log-likelihood}
\end{align}

\noindent
{\bf [\,Smoothing\,]}

The smoothing density is obtained recursively by
\begin{align}
 s_i \: =\: \tilde s(x_i) 
 = \tilde f(x_i)\sum_{j=1}^{k} \int_{x_{j-1}}^{x_j} \frac{\tilde q(x_i-u) \tilde s(u)}{\tilde p(u)} du 
 = \Delta x \cdot f_i \sum_{j=1}^{k} \frac{q_{i-j} s_{j}}{p_{j}}. \label{Eq_NGF_smoothing_1D}
\end{align}

In practical implementations, in order to make the algorithm robust, normalize the density functions so that the integral over the entire domain equals 1, each time after performing the prediction step (\ref{Eq_NGF_prediction_1D}) and the smoothing step (\ref{Eq_NGF_smoothing_1D}). 

For example, by replacing $f_i$ with $f_i / I(f)$, where $I(f)$ is defined as follows: 
\begin{equation}
 I(f) = \int_{x_0}^{x_k} f(x) dx = \Delta x \left( f_1 + \cdots + f_k \right). \label{ngauss-3-5-1}
\end{equation}

The computational complexity of the above algorithm increases quadratically with the number of grid points in one dimension but grows exponentially with the dimension of the state vector. 
This limitation motivates the adaptive moving-grid strategy described in the following subsection.

\subsection{Adaptive Moving-Grid Method}

In the previous subsection, the computational domain was assumed to be a fixed finite interval for simplicity of presentation.
For many nonstationary state-space models, however, the latent state may drift over a much wider range than the local uncertainty.
Consequently, the posterior density may occupy only a small portion of
the computational domain, which substantially reduces the
accuracy of the numerical approximation.

To overcome this difficulty, we shift the computational grid at every time step.
Let \(i_{\max}\) be the location index at which the filtering density $f_i$ attains its maximum.
The location index is updated recursively according to
\begin{align}
\mathrm{LOC}_{n}=\mathrm{LOC}_{n-1} + i_{\max}.
\end{align}
At each time step, the computational domains for the filtering distribution
$p(x_{n}|Y_{1:n})$ and the predictive distribution $p(x_{n+1}|Y_{1:n})$ are shifted to
\begin{align}
[x_0+\mathrm{LOC}_{n}\Delta x,\;
 x_k+\mathrm{LOC}_{n}\Delta x].
\end{align}

By adaptively translating the computational domain in this manner, 
the region of high probability density remains near the center of the computational grid.
Consequently, even when the trend exhibits large long-term drifts,
the posterior distributions can be approximated accurately using 
a relatively small number of grid points.
This adaptive moving-grid strategy substantially improves numerical
accuracy without increasing the number of grid points.
An example illustrating the evolution of the location index $\mathrm{LOC}_{n}$ is presented in Section 3.4.

The adaptive moving-grid method is particularly effective for nonstationary time series whose state trajectories exhibit large long-term variations.

\section{Maximum Likelihood Estimation of State-Space Model Parameters}
This section demonstrates the advantages of the non-Gaussian filter for maximum likelihood estimation by comparing its deterministic likelihood with the stochastic likelihood obtained by the particle filter.
\subsection{Log-Likelihood Computation}

Let $\theta$ denote the vector of unknown parameters in the state-space model.
The log-likelihood function is defined as
\begin{align}
  \ell(\theta)  =  \log p(Y_{1:N})
  =  \sum_{n=1}^{N}  \log p(y_n|Y_{1:n-1}).
\end{align}
For the particle filter, the predictive density can be approximated by
\begin{align}
  \hat{p}(y_n|Y_{1:n-1})  =  \frac{1}{m} \sum_{j=1}^{m} p(y_n|p_n^{(j)})
    = \frac{1}{m} \sum_{j=1}^{m} \alpha_n^{(j)},
  \label{EQ_PF_log-likelihood}
\end{align}
where $p_n^{(1)},\ldots,p_n^{(m)}$ are the particles approximating the predictive distribution at time $n$.
Accordingly, the log-likelihood is approximated by
\begin{align}
  \hat{\ell}(\theta)
  = \sum_{n=1}^{N} \log \biggl( \frac{1}{m} \sum_{j=1}^{m} \alpha_n^{(j)} \biggr).
\end{align}

Although the particle filter provides an unbiased estimator of the
likelihood, the resulting estimator of the log-likelihood is
negatively biased due to Jensen's inequality. It has been shown that, 
under the asymptotic situation in which $N$, $m \rightarrow \infty$ with $N/m$ fixed, 
the bias is evaluated as
\begin{align}
  \mathrm{Bias}(\hat{\ell})  \approx  -\frac{1}{2}\frac{NV}{m},
\end{align}
where $V$ denotes the asymptotic variance rate of the log-likelihood
estimator, $m$ is the number of particles and $N$ is the number of observations (B\'{e}rard et al., 2014).

Although the bias of the log-likelihood can be corrected using this asymptotic bias approximation formula, the variance of the estimated log-likelihood remains the principal obstacle to maximum likelihood estimation, as shown in the examples below.

\subsection{Example: Simple Trend Model}
\begin{figure}[bp]
\begin{center}
\includegraphics[width=100mm,angle=0,clip=]{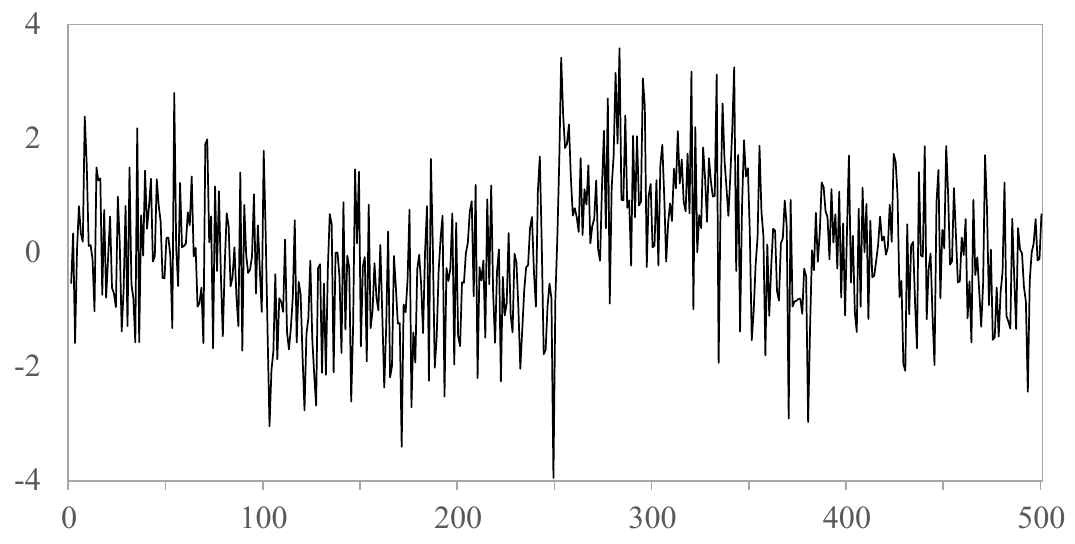}
\end{center}
\caption{Test data generated by the model of (\ref{Eq_test-data_with_jumps}).}
\label{Fig_test-data_with_jumps}
\end{figure}

To illustrate the behavior of the likelihood function,
we first consider the following synthetic data set.
Figure \ref{Fig_test-data_with_jumps} shows the test data generated from the following model
(Kitagawa, 1987, 1996);
\begin{align}
 y_n \sim \mathcal{N}(\mu_n,1), \label{Eq_test-data_with_jumps}
\end{align}
where \( \mu_n = 0\) for \( 1\leq n \leq 100 \), 
  1, for \( 101\leq n \leq 250 \),
 \(-1\), for \( 251\leq n \leq 350 \) and
  0, for \( 351\leq n \leq 500 \).
The series exhibits three abrupt level shifts. 

To estimate the underlying piecewise-constant mean, shown in Figure \ref{Fig_test-data_with_jumps}, we consider the following first-order trend model; 
\begin{align}
 x_n &= x_{n-1} + v_n \nonumber \\
 y_n &= x_n \:+\: w_n, \label{Eq: trend_model}
\end{align}
where $x_n$, $v_n$ and $w_n$ represent the unknown trend at time $n$, the system noise and the observation noise, respectively. 
The system noise $v_n$ and the observation noise $w_n$ are assumed to follow the Gaussian distribution $\mathcal{N}(0,\tau^2)$ and $\mathcal{N}(0,\sigma^2)$, respectively.

Although the proposed method is intended for nonlinear and
non-Gaussian state-space models,
we begin with a linear Gaussian model because the Kalman filter provides
the exact posterior distributions and the exact likelihood.
This allows the approximation error of the non-Gaussian filter to be
evaluated quantitatively.
More challenging nonlinear and non-Gaussian models are considered in the following subsections.

The predictive mean and variance, $x_{n|n-1}$ and $V_{n|n-1}$, and the filtered mean and variance,
$x_{n|n}$ and $V_{n|n}$, of the state-space model are obtained by the Kalman filter.
The exact log-likelihood is then given by
\begin{align}
\ell(\theta)
= \log L(\theta)
= \sum_{n=1}^{N} \log p(y_n|Y_{1:n-1})
= -\frac12 \biggl\{N\log(2\pi) + \sum_{n=1}^{N}\log r_n
  + \sum_{n=1}^{N} \frac{\varepsilon_n^2}{r_n} \biggr\},
\end{align}
where $\varepsilon_n = y_n-x_{n|n-1}$,  $r_n = V_{n|n-1} + \sigma_n^2$.

\begin{figure}
\begin{center}
\includegraphics[width=140mm,angle=0,clip=]{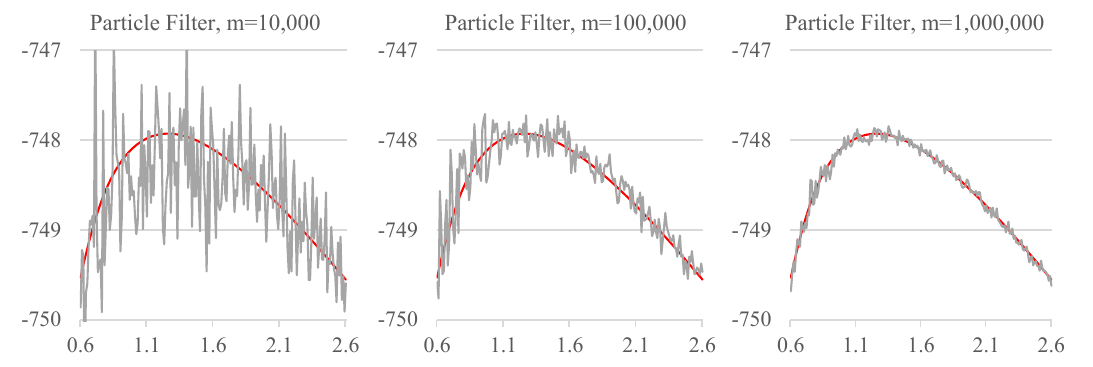}
\includegraphics[width=140mm,angle=0,clip=]{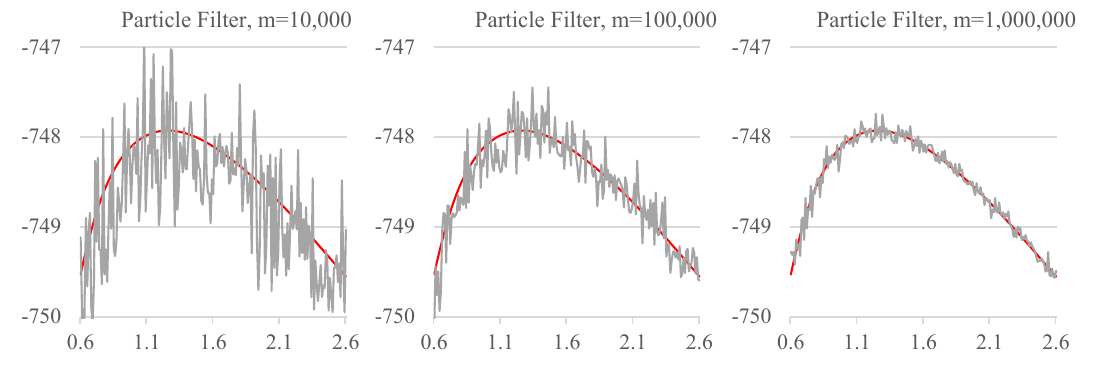}
\end{center}
\caption{Log-likelihood of the Gaussian trend model obtained by particle filter with various $M$. The top row shows the results when resampling is performed only when ESS$ < 0.5$; the bottom row shows the results when resampling is always performed. The red curves represent the exact log-likelihood computed by the Kalman filter. The three columns correspond to $m=10^4$, $10^5$ and $10^6$ particles.}\label{Fig:log-likelihood_particle-filter}
\end{figure}

We first investigate the log-likelihood function obtained by the particle filter.
Figure \ref{Fig:log-likelihood_particle-filter} shows the log-likelihood functions 
computed by the particle filter.
%
%
%
The same sequence of random numbers was used for all parameter values.
Even when the same sequence of random numbers is used, 
the estimated log-likelihood exhibits substantial Monte
Carlo fluctuations even when the number of particles is as large as $m=10^6$.
Consequently, gradient-based optimization becomes unreliable.
It is also seen that adaptive resampling based on
ESS results in slightly smaller fluctuations than resampling at every time step.
In the remainder of this paper, we always apply resampling based on ESS.

For the log-likelihood estimated by the particle filter, it has been shown that
\begin{align}
\log\hat \ell_m-\log \ell_m \xrightarrow[m\rightarrow \infty]{} \mathcal{N}\Bigl( -\frac12\alpha\nu^2, \alpha\nu^2 \Bigr),
\end{align}
where $\alpha = \lim\limits_{m=\infty}N/m$ and $\nu^2$ denotes the asymptotic variance rate per observation (B\'erard et al., 2014).

\begin{figure}
\begin{center}
\includegraphics[width=140mm,angle=0,clip=]{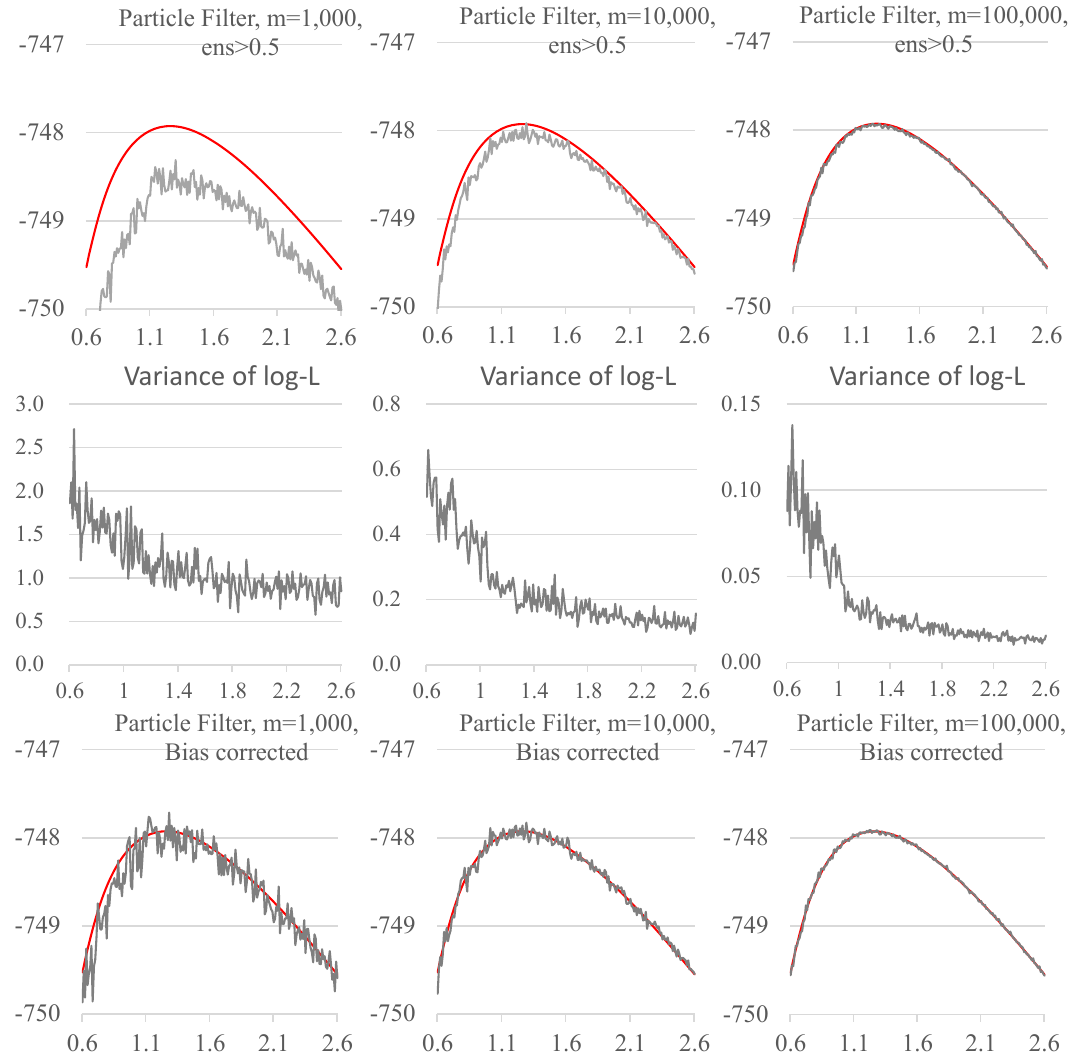}
\end{center}
\caption{Bias and variace of the log-likelihood of the Gaussian trend model. The red curves show the exact log-likelihood and the gray curves show the average of 100 particle filter estimates.
From left to right, the figures correspond to $m=10^3$, $10^4$ and $10^5$ particles. The horizontal axis indicates $\tau^2$.}
\label{Fig: bias-variance of PF log-likelihood}
\end{figure}

This asymptotic result explains both the negative bias and the variance observed in
Figure \ref{Fig: bias-variance of PF log-likelihood}.
Compared with Figure \ref{Fig:log-likelihood_particle-filter}, 
the Monte Carlo variability is greatly reduced, making the negative bias clearly visible.
The middle row shows the estimated variance of $\log\hat L$.
The variance is not constant over $\tau^2$ and decreases as $\tau^2$ increases.
The bottom row shows the bias-corrected log-likelihood obtained by
adding one-half of the estimated variance to the estimated log-likelihood.
The bias is almost removed, which agrees well with the asymptotic theory.

Although repeated runs reduce the Monte Carlo variance,
the reduction is proportional to the inverse square root of the number of repetitions.
Consequently, a substantial computational effort is required to obtain a smooth
likelihood surface.

Possible alternatives for maximizing such noisy likelihoods include obtaining a rough estimate through grid search or employing Bayesian optimization based on Gaussian processes.
While Bayesian optimization is appealing, it is difficult to implement for multidimensional parameters.

\begin{figure}
\begin{center}
\includegraphics[width=140mm,angle=0,clip=]{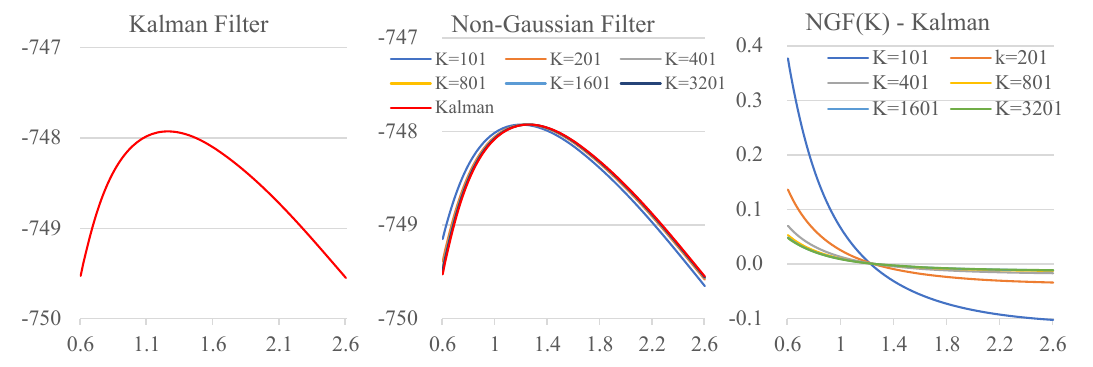}
\end{center}
\caption{Log-likelihood of the Gaussian trend model obtained by Kalman filter and non-Gaussian filter with various $K$. The left panel shows 
the exact log-likelihood function computed by the Kalman filter for
$\theta= 10^2\times\tau^2$ over the range $0.6\le\theta\le2.6$.
The middle panel shows the log-likelihood functions obtained by the
non-Gaussian filter for $k=100$, 200, 400, 800, 1600 and 3200. The right panel 
plots the difference between the
log-likelihoods computed by the non-Gaussian filter and the Kalman filter.}
\label{Fig:true_log-likelihood}
\end{figure}

The non-Gaussian filter avoids these difficulties because it computes the likelihood deterministically through equation (\ref{Eq_NGF_log-likelihood}).
Figure \ref{Fig:true_log-likelihood} shows the log-likelihood functions
computed by the non-Gaussian filter.
Except for the case $k=100$,
the approximated log-likelihood functions are almost indistinguishable
from the exact log-likelihood computed by the Kalman filter (red curve),
indicating that the maximum likelihood estimator can be accurately
obtained from the non-Gaussian filter approximation.
As shown in the right panel, which shows the difference from the true 
log-likelihood obtained by Kalman filter,
the approximation exhibits a positive bias for $\theta<1.27$
and a negative bias for $\theta>1.27$.
The bias decreases rapidly as the number of grid points $k$ increases.
The approximated likelihood converges rapidly to the exact likelihood as
the number of grid points increases.

\begin{table}[bp]
\caption{Maximum likelihood estimates of $\tau^2$, the maximum log-likelihoods of nonlinear state-space model and the cpu-time in second obtained by non-Gaussian filter with various $k$ (Left) and particle filter with various $m$ (right).}\label{Tab:MLE_nonlinear-model}\label{Tab:MLE_Gauss}
\begin{center}
\medskip
{ \tabcolsep = 5pt
\begin{footnotesize}
\begin{tabular}{c|ccc||c|ccc} \hline 
     \multicolumn{4}{c||}{Non-Gaussian Filter} & \multicolumn{4}{c}{Particle Filter} \\ \hline
 $k$ & $\hat{\sigma}^2$ & $\ell (\hat{\sigma}^2)$ & CPU time &
 $m$ & $\hat{\sigma}^2$ & $\ell (\hat{\sigma}^2)$ & CPU time \rule[-3pt]{0pt}{14pt}\\ \hline 
  50 & 0.009918 & $-747.1646$ &  0.01 & 1000    & 0.010259 (0.003607) & $-744.2565$ (0.7019) & 0.13\rule[-2pt]{0pt}{14pt} \\
 100 & 0.010165 & $-747.9234$ &  0.02 & 10000   & 0.012408 (0.001409) & $-746.0530$ (0.1402) & 1.31\\
 200 & 0.011602 & $-747.9202$ &  0.05 & 100000  & 0.011303 (0.001405) & $-747.1046$ (0.1368) & 13.69\\
 400 & 0.012146 & $-747.9192$ &  0.16 & 1000000 & 0.012257 (0.000942) & $-747.7129$ (0.0187) & 137.00\\ \cline{5-8}
 800 & 0.012243 & $-747.9191$ &  0.58 &  \multicolumn{4}{c}{ } \\
1600 & 0.012268 & $-747.9190$ &  2.34 &  \multicolumn{4}{c}{ }\\
3200 & 0.012274 & $-747.9190$ &  8.59 &  \multicolumn{4}{c}{ }\\
6400 & 0.012276 & $-747.9190$ & 34.17 &  \multicolumn{4}{c}{ }\rule[-2pt]{0pt}{10pt} \\ \hline
\end{tabular}
\end{footnotesize}
}
\end{center}
\end{table}

Table~\ref{Tab:MLE_Gauss} summarizes the behavior of the maximum likelihood estimates
obtained by the non-Gaussian filter and the particle filter.
It can be seen that, the estimates by the non-Gaussian filter stabilize rapidly for \(k \ge 400\), and the maximum log-likelihood values remain almost constant.
As the grid size doubles, the CPU time increases fourfold.

The right half of Table~\ref{Tab:MLE_Gauss} summarizes the results by the particle filter.
For each particle size, the optimization procedure was repeated ten times using different random number sequences.
In each run, the maximum log-likelihood and the corresponding parameter
value were obtained by searching approximately 1,000 grid points over the parameter space.
The table reports the averages of the resulting maximum likelihood
estimates and maximum log-likelihood values.
The numbers in parentheses denote the corresponding standard deviations.

The resulting estimates exhibit substantial variability.
For instance, the approximate 95\% confidence interval
($\pm 1.96$ standard deviations) for the parameter estimate is $0.00855<\theta<0.01406$
when $m=10,000$.
Even with as many as $m=1,000,000$ particles, the confidence interval remains relatively wide,
$0.01041<\theta<0.01410$.
This indicates that the Monte Carlo variability of the likelihood estimate
remains a serious obstacle to maximum likelihood estimation.

Table~\ref{Tab:Change_of_domain} demonstrates the importance of selecting an appropriate state
domain for the non-Gaussian filter.
The second through fifth columns compare the posterior distributions of
the state (both the filtering and smoothing distributions), obtained
using the maximum likelihood estimate of the parameter for each value of
$k$, with the corresponding posterior distributions computed by the
Kalman filter, which are regarded as the reference solution.
The second and third columns show the results obtained with the state
domain $-8 < x < 8$, whereas the fourth and fifth columns correspond to
the narrower domain $-4 < x < 4$.

For both the filtering and smoothing distributions, the errors decrease
rapidly as the number of grid points $k$ increases, indicating that the
accuracy of the non-Gaussian filter improves with finer discretization.
Interestingly, when $k \le 200$, the narrower domain
$-4 < x < 4$ yields more accurate results than the wider domain $-8 < x < 8$.
However, for $k \ge 800$, virtually no further improvement is observed
for the narrower domain.

The sixth through ninth columns evaluate the estimation accuracy of the
posterior mean of the state rather than that of the posterior distribution itself.
The accuracy is measured by the mean squared error of the posterior
mean.
Essentially the same tendency is observed.
For the wider domain $-8 < x < 8$, however, the mean squared error
decreases even more rapidly than the error of the posterior
distribution.

These results show that the choice of the computational domain is at least as important as the number of grid points. An excessively wide domain wastes computational resources, whereas an overly narrow domain truncates the posterior distribution and limits the achievable accuracy regardless of the grid resolution.

\begin{table}[tbp]
\caption{Maximum likelihood estimates of $\tau^2$ and the maximum log-likelihoods of Gaussian and Cauchy distribution models obtained by non-Gaussian filter with various $k$.}\label{Tab:Change_of_domain}
\begin{center}
\medskip
\begin{footnotesize}
\begin{tabular}{c|cc|cc|cc|cc} \hline 
    & \multicolumn{4}{c|}{Accuracy of distribution}& \multicolumn{4}{c}{Accuracy of mean} \\
$k$ & \multicolumn{2}{c|}{$[-8:8]$}& \multicolumn{2}{c|}{$[-4:4]$}&  \multicolumn{2}{c|}{$[-8:8]$} & \multicolumn{2}{c}{$[-4:4]$}\\
    &  Filter & Smoother &  Filter & Smoother &  Filter & Smoother & Filter & Smoother\rule[-5pt]{0pt}{16pt} \\ \hline 
  50 & 0.59515 & 0.77605 & 0.17682 & 0.16533 & 0.42498 & 0.27626 & 0.13422 & 0.07076\rule[-2pt]{0pt}{14pt} \\
 100 & 0.20420 & 0.18423 & 0.00792 & 0.00849 & 0.16209 & 0.08558 & 0.00439 & 0.00229 \\
 200 & 0.01526 & 0.01373 & 0.00145 & 0.00187 & 0.01146 & 0.00605 & 0.00023 & 0.00015 \\
 400 & 0.00194 & 0.00212 & 0.00252 & 0.00275 & 0.00074 & 0.00039 & 0.00127 & 0.00071 \\
 800 & 0.00097 & 0.00128 & 0.00293 & 0.00307 & 0.00005 & 0.00003 & 0.00167 & 0.00092 \\
1600 & 0.00093 & 0.00126 & 0.00271 & 0.00271 & 0.00000 & 0.00000 & 0.00178 & 0.00098 \\
3200 & 0.00079 & 0.00108 & 0.00232 & 0.00211 & 0.00000 & 0.00000 & 0.00181 & 0.00100 \\
6400 & 0.00011 & 0.00016 & 0.00271 & 0.00275 & 0.00000 & 0.00000 & 0.00182 & 0.00100 \rule[-4pt]{0pt}{12pt}
 \\ \hline
\end{tabular}\end{footnotesize}
\end{center}
\end{table}

These experiments clearly demonstrate that, for low-dimensional
state-space models, the deterministic likelihood computed by the
non-Gaussian filter provides a much more reliable basis for maximum
likelihood estimation than the stochastic likelihood obtained by the
particle filter.

The results also indicate that, for low-dimensional state-space models, increasing the number of grid points is considerably more effective than increasing the number of particles in the particle filter.

\subsection{Maximum Likelihood Estimation for a Nonlinear State-Space Model}
In this subsection, we investigate whether the proposed maximum likelihood estimation method remains effective for the following nonlinear state-space model:
\begin{align}
  x_n &= \frac{1}{2}x_{n-1} + \frac{25x_{n-1}}{x_{n-1}^2+1} + 8\cos (1.2n) + v_n \nonumber \\
  y_n &= \frac{x_n^2}{10} + w_n, \label{Eq_nonlinear_model}
\end{align}
where $v_n\sim\mathcal{N}(0,1)$, $w_n\sim\mathcal{N}(0,\sigma^2)$,
and $x_0\sim\mathcal{N}(0,5)$ (Kitagawa, 1991, 2020; Gordon et.~al., 1993).
This model has become a standard benchmark for nonlinear filtering and
smoothing because it contains both nonlinear state dynamics and a
nonlinear observation equation while remaining simple enough for
systematic comparison.

\begin{figure}[h]
\begin{center}
\includegraphics[width=140mm,angle=0,clip=]{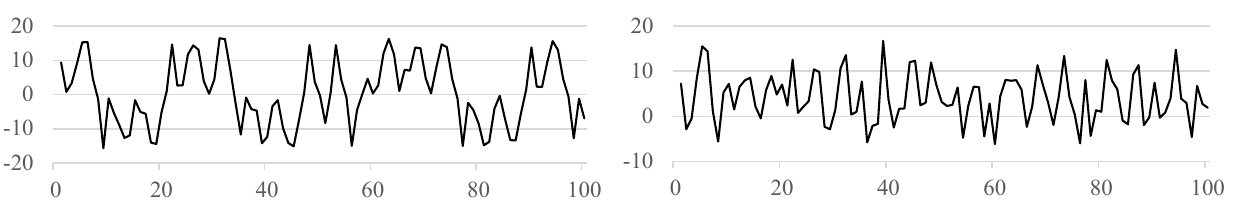}
\end{center}
\caption{Test data for nonlinear smoothing obtained by model (\ref{Eq_nonlinear_model}).
Left: signal $x_n$, right: observed time series $y_n$.}
\label{Fig:nlsim1_data}
\end{figure}

Figure~\ref{Fig:nlsim1_data} presents a simulated data set of length
$N=100$ generated from model (\ref{Eq_nonlinear_model})
with observation noise variance $\sigma^2=10$.
The left panel shows the latent signal, $x_1,\ldots ,x_N$,
while the right panel shows the corresponding observations, $y_1,\ldots ,y_N$.
The same data set is used throughout this subsection to evaluate the
performance of the non-Gaussian filter.

The objective is to estimate the latent state sequence together with the
unknown observation noise variance
from the observations $y_1,\ldots,y_N$.
Throughout this example, the observation noise variance $\sigma^2$
is unknown and is estimated by maximum likelihood.
The log-likelihood function $\ell(\sigma^2)$
is evaluated using the non-Gaussian filter.

\begin{figure}[h]
\begin{center}
\includegraphics[width=140mm,angle=0,clip=]{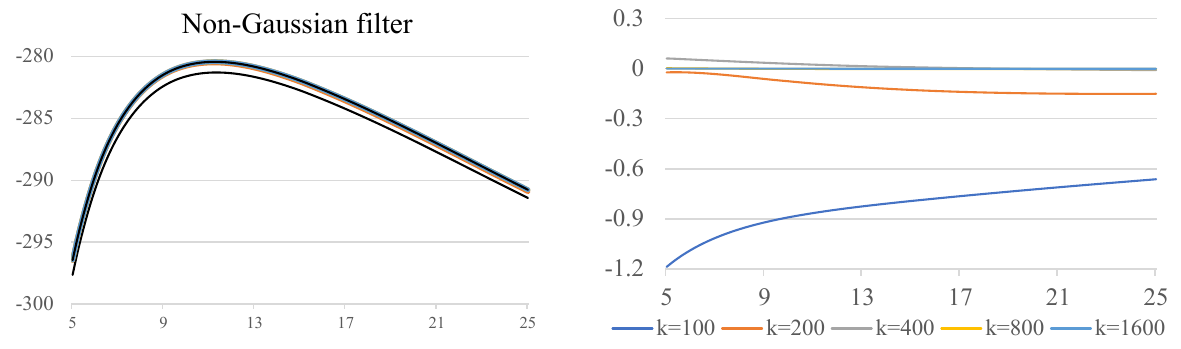}
\end{center}
\caption{Log-likelihood function obtained by the non-Gaussian filter.}
\label{Fig:log-likelihood_NGF}
\end{figure}

The left panel of Figure~\ref{Fig:log-likelihood_NGF} shows the log-likelihood
functions computed with $k=100\times2^j$, $j=0,\ldots,5$,
grid points while varying $\sigma^2$ from 5 to 25.
Except for the coarsest discretization ($k=100$), the curves are almost indistinguishable.

To highlight the effect of the grid resolution more clearly,
the right panel plots the difference between each log-likelihood curve
and the reference curve obtained with $k=3200$.
For $k=100$, the maximum difference exceeds 0.7, whereas for $k=200$
the bias is approximately 0.2.
For $k\ge400$, the approximation error becomes negligible over the entire parameter range.
More importantly, even when $k=100$,
the location of the maximum likelihood estimate is already very close to
that obtained with $k=3200$.
This suggests that reasonably accurate maximum likelihood estimates can
be obtained even with a relatively small number of grid points such as
$k=100$ or 200.

\begin{figure}
\begin{center}
\includegraphics[width=140mm,angle=0,clip=]{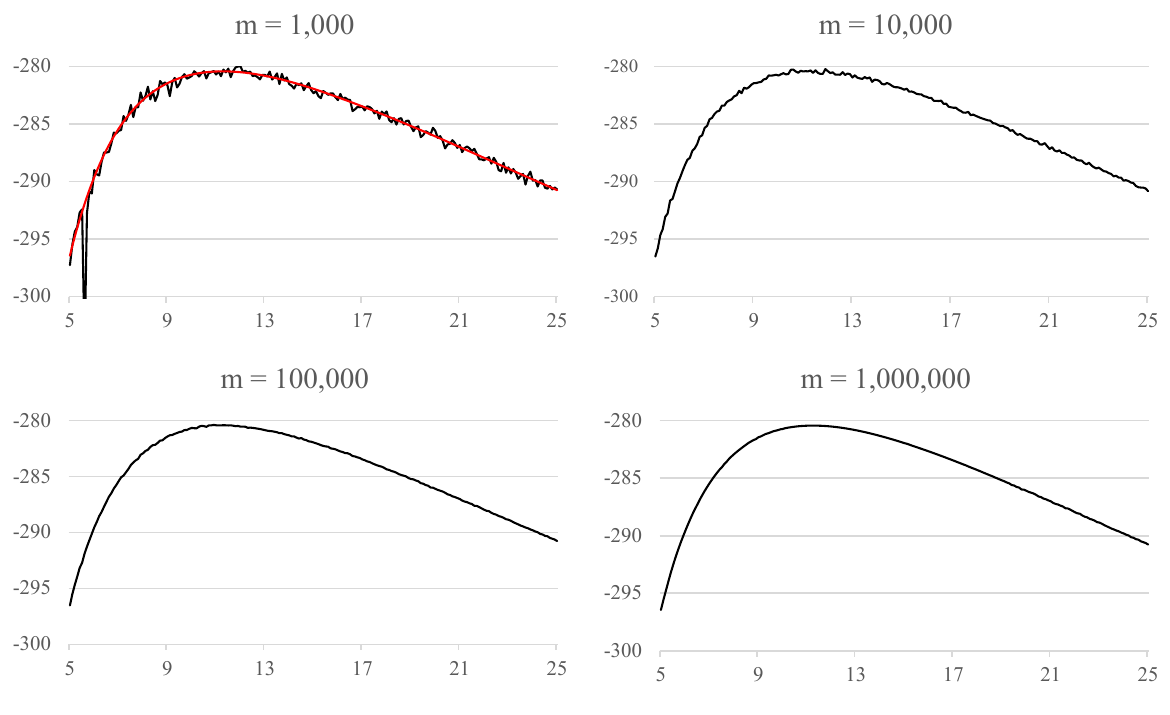}
\end{center}
\caption{Log-likelihood function obtained by the particle filter.}
\label{Fig:log-likelihood_of_MCF_trend-model}
\end{figure}

For comparison, Figure \ref{Fig:log-likelihood_of_MCF_trend-model}
shows the log-likelihood functions obtained by the particle filter.
In this case, compared to the previous example, 
the fluctuation in log-likelihood is smaller, yielding a relatively smooth function.

Table~\ref{Tab:MLE_nonlinear-model} summarizes the convergence of the maximum likelihood
estimates of $\sigma^2$, the corresponding maximum log-likelihood values,
and the computation time per function evaluation when varying the number of grid points \(k\) or particles \(m\).
Since the results by the particle filter depend on the random-number seed, 
the mean and the standard deviation of the maximum likelihood estimates and the maximum log-likelihoods in 10 runs with different random-number seeds are shown.
The reported CPU time is the time required for a single evaluation of
the log-likelihood function by a desktop computer.

\begin{table}[h]
\caption{Maximum likelihood estimates of $\tau^2$ and the maximum log-likelihoods of nonlinear state-space model obtained by non-Gaussian filter with various $k$ and particle filter with various $m$. The first 4 columns show the results for the non-Gaussian filter, and 
the rest show the results by the particle filter.}\label{Tab:MLE_nonlinear-model}
\begin{center}
\medskip
{ \tabcolsep = 5pt
\begin{footnotesize}
\begin{tabular}{c|ccc||c|ccc} \hline 
     \multicolumn{4}{c||}{Non-Gaussian filter} & \multicolumn{4}{c}{Particle filter} \\ \hline
 $k$ & $\hat{\sigma}^2$ & $\ell (\hat{\sigma}^2)$ & CPU time &
 $m$ & $\hat{\sigma}^2$ & $\ell (\hat{\sigma}^2)$ & CPU time \rule[-3pt]{0pt}{14pt}\\ \hline 
 100 & 11.31232 & $-281.2867$ &  0.00 & 1000   & 11.2168 (0.5991) & $-279.9351$ (0.1520) & 0.03 \rule[-2pt]{0pt}{14pt} \\
 200 & 11.20224 & $-280.5194$ &  0.01 & 10000  & 11.2478 (0.3535) & $-280.2478$ (0.0488) & 0.27 \\
 400 & 11.22631 & $-280.4047$ &  0.02 & 100000 & 11.1408 (0.1859) & $-280.3729$ (0.0175) & 2.79 \\
 800 & 11.24098 & $-280.4291$ &  0.09 & 1000000& 10.9149 (0.0733) & $-280.4278$ (0.0091) &27.73 \\ \cline{5-8}
1600 & 11.24140 & $-280.4277$ &  0.33 &  \multicolumn{4}{c}{ }\\
3200 & 11.24151 & $-280.4273$ &  1.27 &  \multicolumn{4}{c}{ }\\
6400 & 11.24153 & $-280.4272$ &  4.95 &  \multicolumn{4}{c}{ }\\
12800& 11.24154 & $-280.4272$ & 19.45 &  \multicolumn{4}{c}{ }\rule[-2pt]{0pt}{10pt} \\ \hline
\end{tabular}
\end{footnotesize}
}
\end{center}
\end{table}


The computational cost increases approximately fourfold when the number
of grid points is doubled, which reflects the quadratic complexity of the
numerical integration.
The maximum likelihood estimates and the maximum log-likelihood values
have essentially converged for $k\ge800$.
Furthermore, since the estimate of $\sigma^2$ obtained with $k=100$
agrees with the converged value to approximately three significant digits,
the non-Gaussian filter provides practically useful parameter estimates
even with a relatively coarse discretization.
The results demonstrate that accurate maximum likelihood estimation can
be achieved without using a fine discretization,
thereby making the non-Gaussian filter computationally practical for
low-dimensional nonlinear state-space models.

The right half of Table~3 summarizes the results obtained by the particle filter.
Although the particle filter yields reasonably accurate estimates on average, 
their variability remains substantial.
Furthermore, the standard deviation of the estimates decreases only
slowly as the number of particles increases.

In contrast, the maximum log-likelihood values converge rapidly to those
obtained by the non-Gaussian filter as the number of particles increases.
Moreover, the standard deviation of the estimated maximum
log-likelihood decreases approximately in proportion to $1/\sqrt{m}$,
which is consistent with the theoretical convergence rate of Monte Carlo estimators.
As shown in Figure \ref{Fig:log-likelihood_of_MCF_trend-model}, 
the variance of the log-likelihood calculated by the particle filter is relatively small in this example.
However, the maximum likelihood estimate of \(\sigma ^{2}\) still exhibits substantial Monte Carlo variability depending on the number of particles \(m\) and the initial seed of the random numbers.

These results confirm that the principal advantage of the non-Gaussian filter 
over the particle filter is preserved even for nonlinear state-space models.
Thus, the deterministic likelihood computed by the non-Gaussian filter
provides a stable foundation for maximum likelihood estimation even in
nonlinear state-space models.

\subsection{Maximum Likelihood Estimation in Radar Tracking Models}
To demonstrate that the proposed method also applies to higher-dimensional state-space models,
we consider standard constant-position (2-D) and constant-velocity (4-D) radar tracking models
(Bar-Shalom et al., 2002; Kitagawa, 1991; Kitagawa and Gersch, 1996).

Let \( (x_n,y_n) \) denote the target position  and let \( (\Delta x_n, \Delta y_n) \) is velocity.  
Define the 2-D and 4-D state vectors by
\begin{equation}
   \boldsymbol{x}_n^{(2)} =
   \left[ \begin{array}{c}
      x_n \\ 
      y_n \\
   \end{array} \right],\quad
   \boldsymbol{x}_n^{(4)} =
   \left[ \begin{array}{c}
      x_n \\ 
      \Delta x_n \\
      y_n \\
      \Delta y_n
   \end{array} \right],
\end{equation}
respectively.
The two-dimensional model represents only the target position, 
whereas the four-dimensional model additionally includes the velocity components.
The state evolution is described by the state equation
\begin{equation}
   \boldsymbol{x}_n^{(\ell)} = F_\ell \boldsymbol{x}_{n-1}^{(\ell)} + G_\ell v_n,
\label{Eq:tracking_system-model}
\end{equation}
where $\ell$ is 2 or 4 and
\begin{align}
  F_2 &= \left[ \begin{array}{cc}
   1 & 0 \\
   0 & 1
 \end{array} \right],
 \quad
 G_2 = \left[ \begin{array}{cc}
   1 & 0 \\
   0 & 1
 \end{array} \right], \nonumber \\
  F_4 &= \left[ \begin{array}{cccc}
   1 & 1 & 0 & 0 \\
   0 & 1 & 0 & 0 \\
   0 & 0 & 1 & 1 \\
   0 & 0 & 0 & 1
 \end{array} \right],
 \quad
 G_4 = \left[ \begin{array}{cc}
   1 & 0 \\
   1 & 0 \\
   0 & 1 \\
   0 & 1
 \end{array} \right],\\
v_n &=\left[\begin{array}{c}
   w_{x,n} \\
   w_{y,n}
\end{array} \right] \sim N(0,\Sigma). \nonumber
\end{align}

The radar measures the range and bearing angle according to
\begin{equation}
   \boldsymbol{y}_n =
   \left[
   \begin{array}{c}
      r_n \\
      \theta_n
   \end{array}
   \right]. \label{Eq:tracking_observation_model}
\end{equation}
The observation model is expressed as
\begin{equation}
   \boldsymbol{y}_n = h(\boldsymbol{x}_n) + \boldsymbol{w_n},
\end{equation}
where
\begin{align}
   r_n & = \sqrt{x_n^2 + y_n^2} + w_{r,n},
\\
   \theta_n &=   \tan^{-1} \left( \frac{y_n}{x_n} \right) + w_{\theta,n}.
\end{align}

\subsubsection{Test Data}

\begin{figure}[tbp]
\begin{center}
\includegraphics[width=110mm,angle=0,clip=]{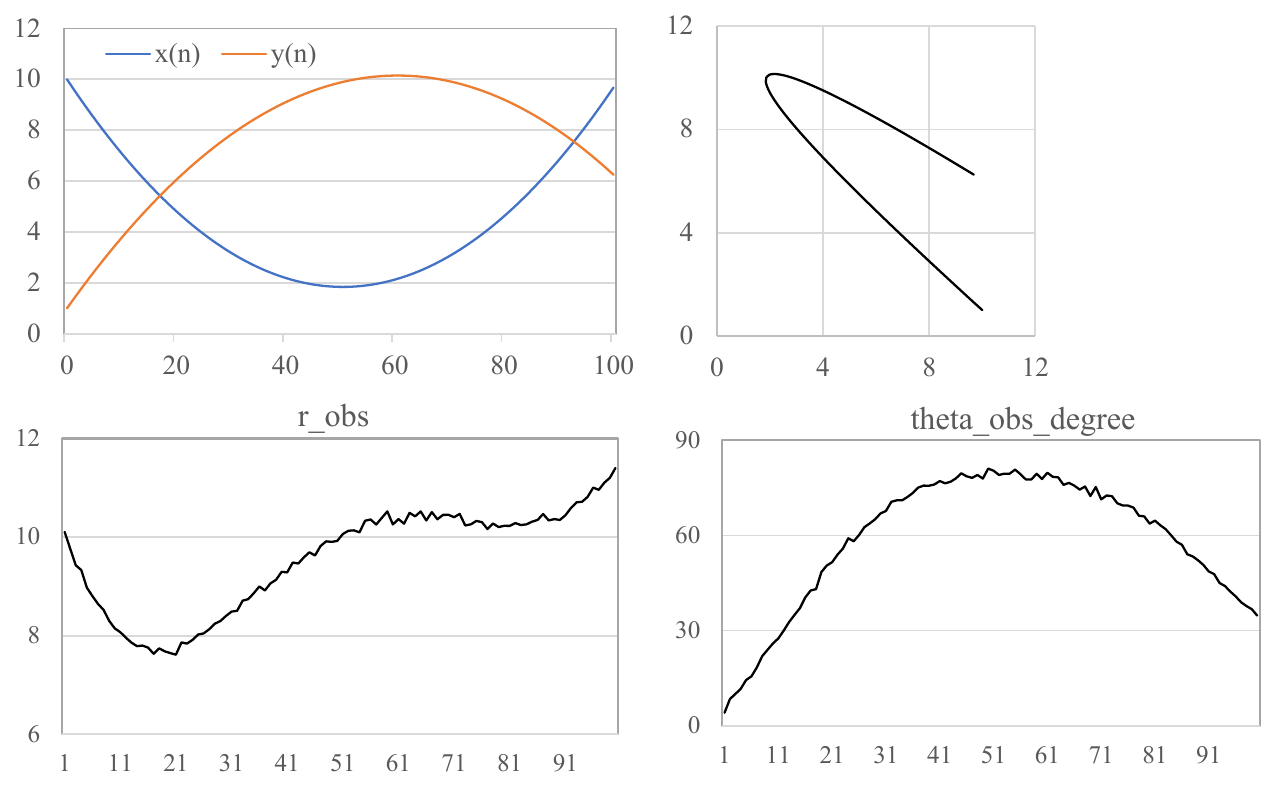}
\end{center}
\caption{Test data of tracking problem.}\label{Fig:tracking_test-data}
\end{figure}

To evaluate the proposed methods, we generated a synthetic data set
that simulates radar observations of a moving target.
Let $(x_n,y_n)$ denote the two-dimensional position of the target.
The true trajectory is assumed to follow the deterministic trajectory given below
\begin{align}
  x_n &= 10 - 0.32n + 0.0064 n(n-1)/2 \nonumber \\
  y_n &=  1 + 0.30n - 0.0050 n(n-1)/2 .
\end{align}
The upper panels of Figure~\ref{Fig:tracking_test-data} shows the generated time series
$x_n$ and $y_n$ for $n=1,\ldots,100$ and
the corresponding trajectory in the two-dimensional plane.

The target is assumed to be observed by a radar that measures the range
$r_n$ and bearing angle $\theta_n$ given by
\begin{align}
  r_n &= \sqrt{x_n^2 + y_n^2} + C u_n, \quad u_n \sim N(0,0.025^2) \nonumber \\
  \theta_n &= \tan^{-1}\left(\frac{y_n}{x_n}\right)\frac{\pi}{180} + C v_n, \quad v_n \sim N(0,0.3^2),
\label{Eq:tracking_observation_model}
\end{align}
where $C$ is a noise amplification factor.
Throughout this example, $C=3$ is used.
The lower panels of Figure~8 show the resulting observations of the
range and bearing angle.

Our objective is to estimate the target position $(x_n,y_n)$
from the observed sequences $(r_n,\theta_n)$.
The observation noise variance is estimated by maximum likelihood.
For this purpose, the log-likelihood function is first evaluated using both the
non-Gaussian filter and the particle filter based on the system model
given by (\ref{Eq:tracking_system-model}) and the observation model 
given by (\ref{Eq:tracking_observation_model}).

\subsubsection{Log-Likelihood}
\begin{figure}[h]
\begin{center}
\includegraphics[width=160mm,angle=0,clip=]{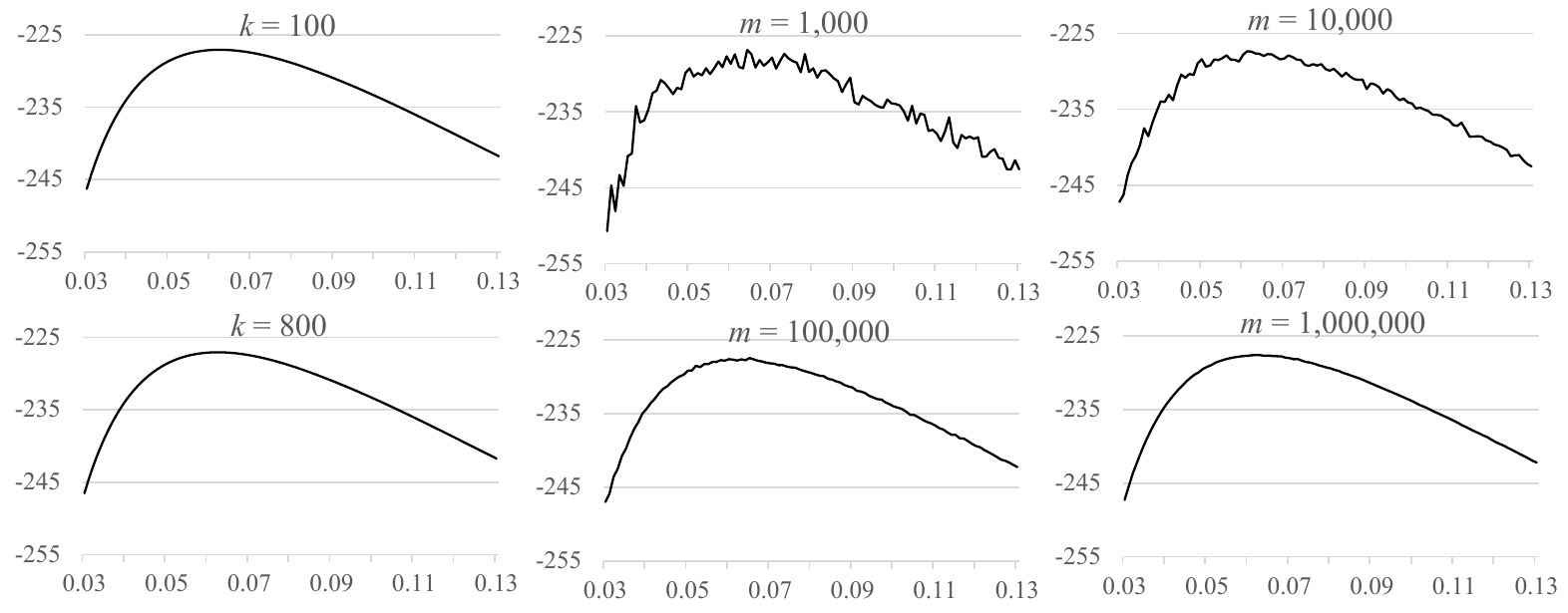}
\end{center}
\caption{Log-likelihood of the 2D-tracking model for test data 1. The left column: obtained by non-Gaussian filter, the right column: obtained by particle filter.}
\label{Fig:2D_tracking_log-likelihood}
\end{figure}

Figure~\ref{Fig:2D_tracking_log-likelihood} compares the log-likelihood functions 
obtained by the two methods.
The left column of Figure~\ref{Fig:2D_tracking_log-likelihood} shows the log-likelihood functions
computed by the non-Gaussian filter for the two-dimensional model using
$k=100$ and $k=800$ grid points.
In both cases the resulting log-likelihood functions are smooth and
continuous, and even the coarse discretization with $k=100$
produces a curve that is almost identical to that obtained with $k=800$.

The two right columns show the corresponding results obtained by the
particle filter using $m=10^k$, $k=3,\ldots,6$, particles.
For $m=10^3$, the log-likelihood surface is highly irregular, making direct maximization difficult.
Only for $m=10^6$ does the log-likelihood become reasonably smooth.

\subsubsection{Maximum Likelihood Estimation}
Table \ref{Tab:Tracking_2D-model} summarizes the maximum likelihood estimates for the
two-dimensional model with $\ell=2$ in (\ref{Eq:tracking_system-model}).
The left half of the table presents the results obtained by the non-Gaussian filter.
Because both spatial coordinates are discretized using the same number of grid points,
the total number of evaluation points is $k\times k$.
Even with $k=100$, the estimates are essentially identical with those obtained using $k=800$
to at least two significant digits.

The right half of the table presents the corresponding particle filter results.
Because the log-likelihood depends on the Monte Carlo random particle realization,
the optimization was repeated ten times using defferent random-number sequences.
The table reports the mean and standard deviation of the resulting estimates.
Although the mean estimates are comparable to those obtained by the
non-Gaussian filter even for relatively small particle numbers,
the standard deviations remain large, indicating that the accuracy of an
individual optimization is limited to approximately one significant digit.

\begin{table}[tbp]
\caption{Maximum likelihood estimates of $\tau^2$ and the maximum log-likelihoods of 2-D state-space model obtained by non-Gaussian filter with various $k$ and particle filter with various $m$.}\label{Tab:Tracking_2D-model}
\begin{center}
\medskip
{ \tabcolsep = 5pt
\begin{footnotesize}
\begin{tabular}{c|ccc||c|ccc} \hline 
     \multicolumn{4}{c||}{Non-Gaussian filter} & \multicolumn{4}{c}{Particle filter} \\ \hline
 $k$ & $\hat{\sigma}^2$ & $\ell (\hat{\sigma}^2)$ & CPU time &
 $m$ & $\hat{\sigma}^2$ & $\ell (\hat{\sigma}^2)$ & CPU time \rule[-3pt]{0pt}{14pt}\\ \hline 
 100 & 0.062196 & $-227.0801$ &  0.44 & 1000   & 0.0642 (0.006989) & $-226.4562$ (0.4592) & 0.05 \rule[-2pt]{0pt}{14pt} \\
 200 & 0.062296 & $-227.0735$ &  2.67 & 10000  & 0.0626 (0.003406) & $-227.3376$ (0.1142) & 0.52 \\
 400 & 0.062321 & $-227.0702$ & 17.89 & 100000 & 0.0630 (0.001491) & $-227.5456$ (0.0773) & 4.94 \\
 800 & 0.062327 & $-227.0685$ &158.36 & 1000000& 0.0628 (0.000750) & $-227.6084$ (0.0111) &49.88 \rule[-2pt]{0pt}{10pt} \\ \hline
\end{tabular}
\end{footnotesize}
}
\end{center}
\end{table}

\begin{figure}[h]
\begin{center}
\includegraphics[width=160mm,angle=0,clip=]{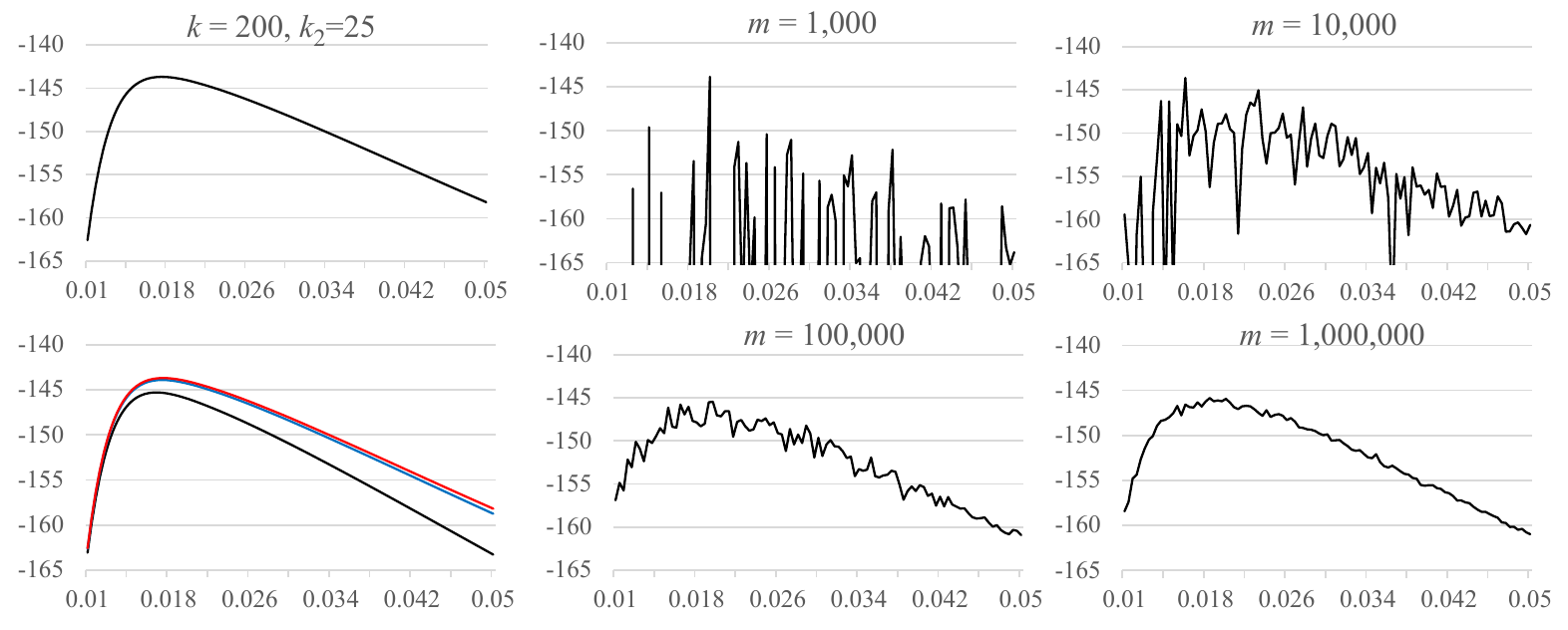}
\end{center}
\caption{Log-likelihood of the 4D-tracking model for test data 1. Left column: obtained by non-Gaussian filter, right column: obtained by particle filter.}\label{Fig:tracking_4D_log-likelihood}
\end{figure}

Figure \ref{Fig:tracking_4D_log-likelihood} presents the corresponding results for the four-dimensional model.
The left column shows the log-likelihood functions obtained by the non-Gaussian filter.
The upper panel corresponds to $k=100$ and $k_2=25$,
where the four-dimensional grid contains $100\times25\times100\times25$ evaluation points.

The lower panel compares the results for $k=50$, 100, and 200, while fixing $k_2=25$.
The log-likelihood obtained with $k=50$ underestimates the true value considerably,
whereas increasing the grid resolution from \(k=100\) to \(k=200\)
changes the likelihood only slightly, indicating numerical convergence.
The two right columns show the particle filter results. 
The contrast with the non-Gaussian filter becomes even more pronounced
for the four-dimensional model.
The likelihood surfaces exhibit much larger Monte Carlo fluctuations.
For $m=10^3$ the fluctuations are too severe for reliable maximum likelihood estimation,
and noticeable irregularity remains even when $m=10^6$ particles are used.

\begin{table}[tbp]
\caption{Maximum likelihood estimates of $\tau^2$ and the maximum log-likelihoods of 4-D state-space model obtained by non-Gaussian filter with various $k$ and particle filter with various $m$.}\label{Tab:Trackibg_4D-model}
\begin{center}
\medskip
{ \tabcolsep = 5pt
\begin{footnotesize}
\begin{tabular}{cc|ccc||c|ccc} \hline 
     \multicolumn{5}{c||}{Non-Gaussian filter} & \multicolumn{4}{c}{Particle filter} \\ \hline
 $k$ & $k_2$ & $\hat{\sigma}^2$ & $\ell (\hat{\sigma}^2)$ & CPU time &
 $m$ & $\hat{\sigma}^2$ & $\ell (\hat{\sigma}^2)$ & CPU time \rule[-3pt]{0pt}{14pt}\\ \hline 
  50 & 25 & 0.016778 & $-145.2944$ &  87.70 & 1000  & 0.02016 (0.003782) & $-145.2554$ (2.3630) & 0.05 \rule[-2pt]{0pt}{14pt} \\
 100 & 25 & 0.017352 & $-143.9129$ & 258.63 & 10000  & 0.01896 (0.002601) & $-144.3904$ (0.9130) & 0.53 \\
 200 & 25 & 0.017450 & $-143.6956$ &1823.31 & 100000 & 0.01864 (0.001830) & $-145.3072$ (0.2876) & 4.92 \\
     &    &   &   &    & 1000000& 0.01820 (0.001151) & $-145.8416$ (0.1588) &49.28 \rule[-2pt]{0pt}{10pt} \\ \hline
\end{tabular}
\end{footnotesize}
}
\end{center}
\end{table}

Table~\ref{Tab:Trackibg_4D-model} compares the maximum likelihood estimates obtained by the
non-Gaussian filter and the particle filter for the four-dimensional model.
The non-Gaussian filter provides stable parameter estimates already with
$k=100$ and $k_2=25$, although the four-dimensional numerical integration requires
considerable computational effort because numerical integration must be performed over a
four-dimensional grid.

In contrast, the particle filter requires much less computation for a single log-likelihood evaluation.
However, its optimization requires many more function evaluations because of the
noisy log-likelihood surface,
and the resulting estimates still exhibit substantial variability due to Monte Carlo randomness.

\subsubsection{State Estimation}
\begin{figure}[h]
\begin{center}
\includegraphics[width=120mm,angle=0,clip=]{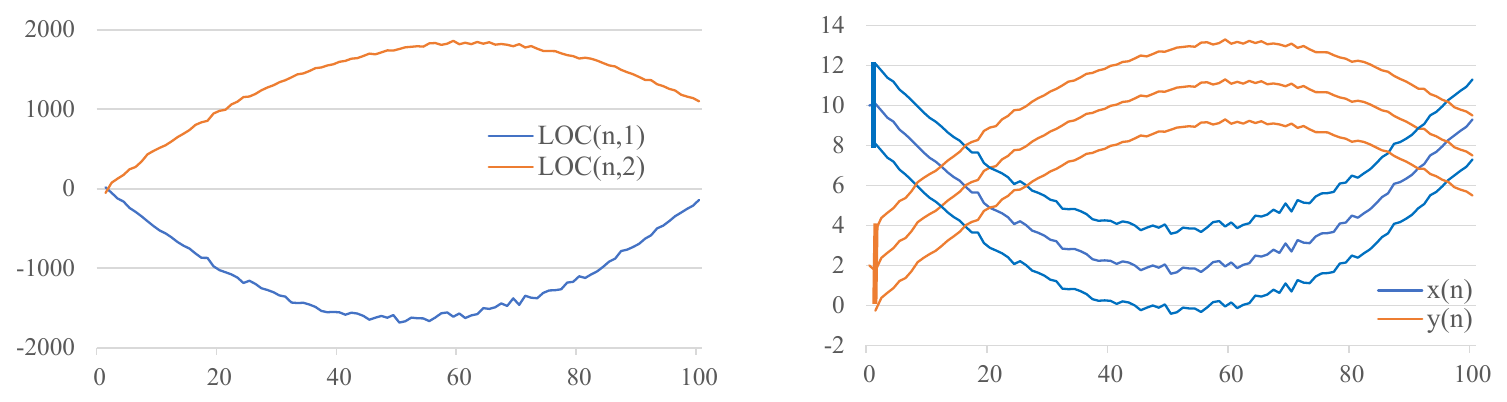}
\end{center}
\caption{Location index and adaptive moving-grid for 2-D model with $k=800$. Left: location indices, LOC(n,1) and LOC(n,2), Right: center of moving-grid. Two the thick vertical lines at the left end of the plot indicate the initial position of the moving-grid at time \(n=0\).}\label{Fig:tracking_LOC_2D-model}
\label{Fig:tracking_location-index}
\end{figure}
The left panel of Figure \ref{Fig:tracking_location-index} shows an example of the location indices, LOC\((n,j)\), \(j=1,2\), for the 2-D model. 
Although both indices start from 0 at \(n=0\), they vary relatively smoothly over time according to the adaptive moving-grid update as the object moves.
The initial moving grid is shown at the left end of the right figure. 
The three curves starting from the initial moving grid indicate the center, maximum and minimum positions of the moving grid in the actual state-space.
The center of the moving grid is determined by the location index.
Without the adaptive moving-grid strategy, a wide finite interval would be required, such as \([-2, 14]\). 
In contrast, using the moving grid allows for a smaller initial interval, \([8,12]\) and \([0,4]\) in this case, thereby maintaining high numerical accuracy while using a much smaller grid.

\begin{figure}
\begin{center}
\includegraphics[width=83mm,angle=0,clip=]{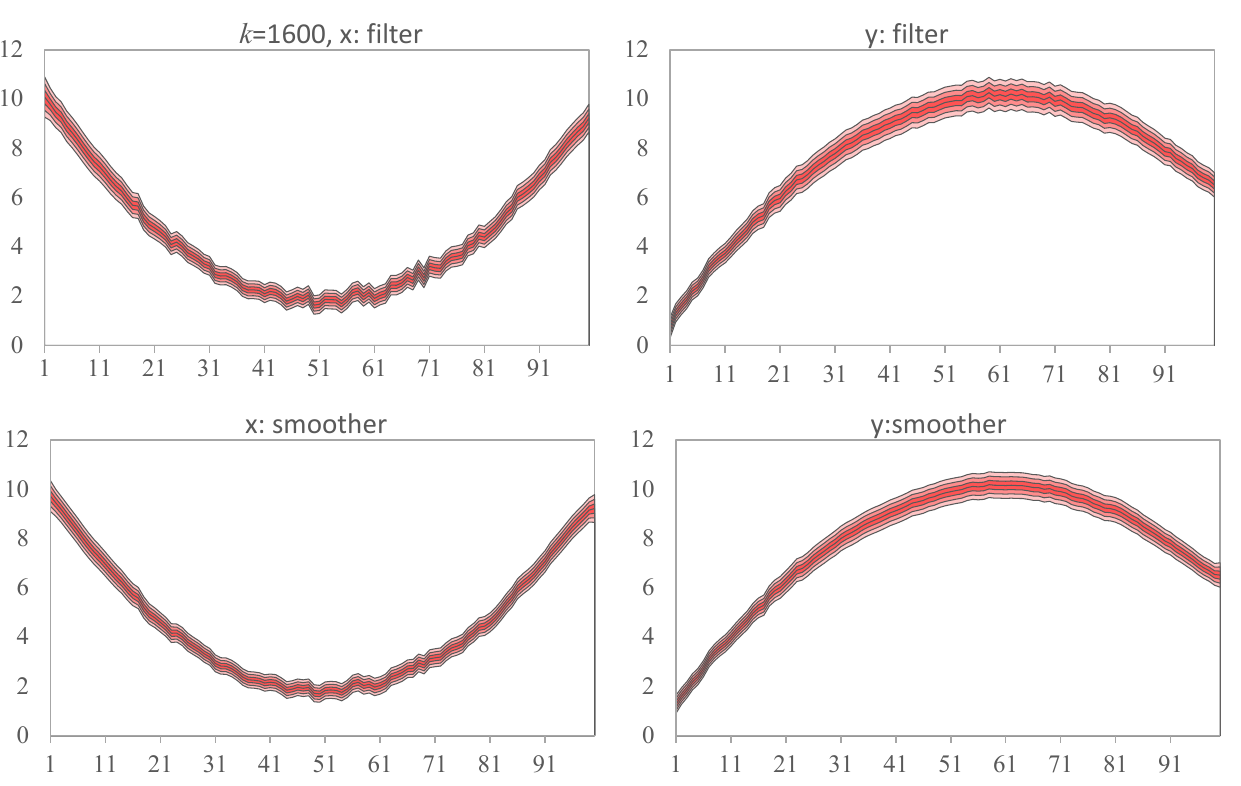}
\includegraphics[width=85mm,angle=0,clip=]{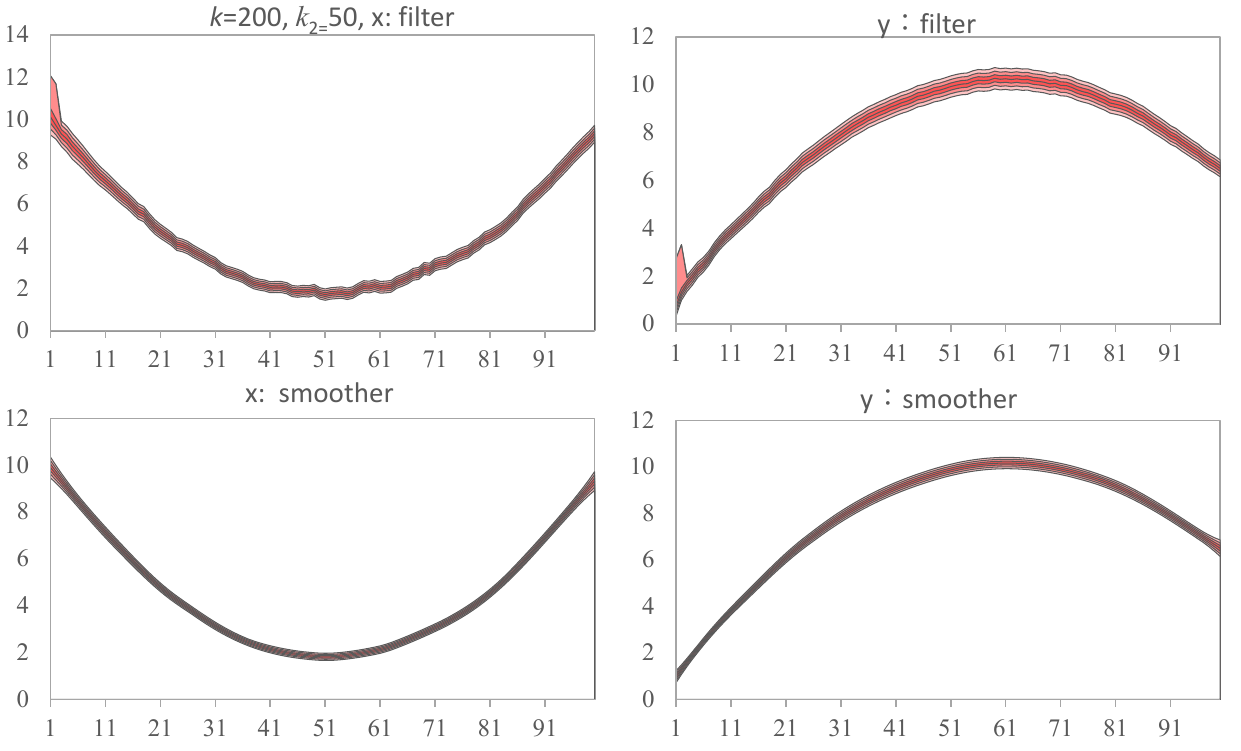}
\end{center}
\caption{Posterior distribution of $x_n$ and $y_n$ by 2D and 4D model. From left to right, $x_n$ and $y_n$ for 2D model, and $x_n$ and $y_n$ for 4D model. Upper row: filtered distribution, lower row: smoothed distribution.}\label{Fig:Tracking_estimated_trace}
\end{figure}

Figure~\ref{Fig:Tracking_estimated_trace} shows the posterior distributions of
$x_n$ and $y_n$ obtained using the maximum likelihood estimates.
Compared with the filtering distributions,
the smoothing distributions evolve more smoothly over time and exhibit
substantially smaller posterior uncertainty.
The posterior distribution of the 4-D model is smoother and more concentrated around the mean compared to that of the 2-D model. 
Although the 2-D model has the advantage of shorter computation time, the 4-D model provides more accurate state estimates.
The computational cost of the non-Gaussian smoother increases rapidly with the state dimension. 
These results suggest a practical hybrid strategy: estimate the model parameters by maximizing the deterministic likelihood computed by the non-Gaussian filter, and then perform state estimation using a particle filter when the state dimension becomes too large for deterministic numerical integration..

\begin{figure}
\begin{center}
\includegraphics[width=165mm,angle=0,clip=]{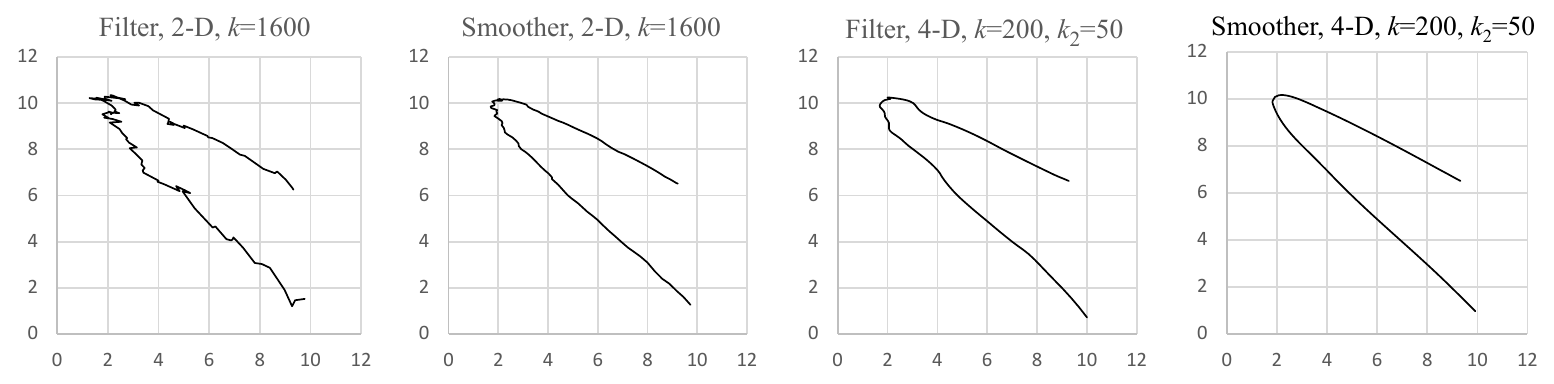}
\end{center}
\caption{Estimated traces on x-y space obtained by 2D filter, 2D smoother, 4D filter and 4D smoother.}\label{Fig:tracking_posterior_mean_trajectory}
\end{figure}

Figure~\ref{Fig:tracking_posterior_mean_trajectory} plots the posterior mean trajectories in the
two-dimensional plane.
For the two-dimensional model, the filtering estimates contain noticeable errors,
whereas smoothing produces a much smoother trajectory.
For the four-dimensional model, the filtering estimates already follow the true trajectory 
reasonably well, although small biases remain.
The smoothing estimates further reduce these errors and yield a very smooth trajectory.
The smoothing estimates are almost indistinguishable from the true trajectory.

This tracking example clearly demonstrates the complementary
characteristics of the two approaches.
The particle filter scales well to higher-dimensional state-space
models but produces a stochastic log-likelihood surface that complicates
maximum likelihood estimation.
In contrast, the non-Gaussian filter yields a deterministic and smooth
likelihood function, making reliable maximum likelihood estimation
possible even for nonlinear observation models.

These results demonstrate that the two methods should be viewed as complementary 
rather than competing approaches.
The present example also suggests that deterministic likelihood
evaluation by the non-Gaussian filter remains practical beyond
one-dimensional models and is applicable to realistic tracking
problems of moderate dimension, whereas the particle filter remains 
the method of choice for state estimation in higher-dimensional systems.

\section{Bayesian Estimation using Self-Organizing State-Space Models}
Whereas the previous section estimated the unknown parameters by maximum
likelihood, this section adopts a fully Bayesian approach in which the
unknown parameters are incorporated into the state vector and estimated
jointly with the latent state. This formulation is known as the
self-organizing state-space model (Kitagawa, 1998).

Maximum likelihood estimation provides point estimates of the unknown
parameters, whereas Bayesian estimation characterizes their uncertainty
through posterior distributions. In addition, maximum likelihood
estimation generally requires repeated evaluations of the log-likelihood
function during numerical optimization. In the self-organizing
state-space model, by contrast, the posterior distributions of both the
parameters and the latent states can be obtained through a single
filtering and smoothing procedure. However, because augmenting the state
vector with unknown parameters increases its dimension, the relative
computational efficiency of the two approaches depends on the model
structure and the number of unknown parameters.

\subsection{Self-Organizing State-Space Modeling for Bayesian Estimation}

Consider the following state-space model:
\begin{align}
  x_n &= f(x_{n-1}) + g(v_n), \qquad v_n \sim \mathcal{P}(0,\tau_n^2,b), \nonumber \\
  y_n &= h(x_n) + w_n, \hspace{8mm}  \qquad w_n \sim \mathcal{N}(0,\sigma_n^2).
\end{align}
Here, $\mathcal{N}(0,\sigma_n^2)$ denotes the Gaussian distribution with
mean zero and variance $\sigma_n^2$, and
$\mathcal{P}(0,\tau_n^2,b)$ denotes the Pearson type VII distribution
with location parameter zero, scale parameter $\tau_n^2$, and shape
parameter $b$. Its probability density function is given by
\begin{align}
  p(x\mid\tau^2,b)
  = \frac{C}{(x^2+\tau^2)^b},
  \qquad -\infty<x<\infty,
  \qquad b>\frac{1}{2},
\end{align}
where
\begin{align}
  C
  = \frac{\tau^{2b-1}\Gamma(b)}
  {\Gamma\left(b-\frac{1}{2}\right)
   \Gamma\left(\frac{1}{2}\right)},
\end{align}
and $\Gamma(\cdot)$ denotes the gamma function.
This family includes the Cauchy and $t$ distributions as
special cases and approaches the Gaussian distribution under an
appropriate limiting parameterization.

Let $\theta_n$ denote the parameter vector specifying the functions
$f$, $g$ and $h$, as well as the distributional parameters
$\tau_n^2$, $b$ and $\sigma_n^2$. The parameter vector is assumed to
follow the random-walk model
\begin{align}
  \theta_n
  = \theta_{n-1}+u_n,
  \qquad
  u_n\sim\mathcal{N}(0,\zeta^2 I).
\end{align}
For time-invariant parameters, that is, when
$\theta_n\equiv\theta$, we set $\zeta^2=0$.

Define the augmented state vector
\begin{align}
  z_n  =
  \begin{bmatrix}
    x_n \\
    \theta_n
  \end{bmatrix}.
\end{align}
The original model can then be rewritten as a state-space model for the
augmented state vector:
\begin{align}
  z_n &= f^*(z_{n-1}) + g^*(v_n,u_n), \nonumber \\
  y_n &= h^*(z_n) + w_n,
\end{align}
where
\begin{align}
  f^*(z_{n-1})
  &=
  \begin{bmatrix}
    f(x_{n-1}) \\
    \theta_{n-1}
  \end{bmatrix},
  &
  g^*(v_n,u_n)
  &=
  \begin{bmatrix}
    g(v_n) \\
    u_n
  \end{bmatrix},
  &
  h^*(z_n)
  &= h(x_n).
\end{align}

Thus, state estimation and parameter estimation are unified within a
single Bayesian filtering and smoothing framework. The marginal
posterior distribution of $\theta_n$ provides the Bayesian parameter
estimate, while that of $x_n$ provides the corresponding state estimate.

\subsection{Example: Single-Parameter Case}

To facilitate comparison with the maximum likelihood approach, we
consider the same trend model as that defined in
(\ref{Eq: trend_model}). Before examining simultaneous state and
parameter estimation, we first present the posterior distributions of
the trend obtained under the Gaussian and Cauchy system-noise models,
with their parameters fixed at the corresponding maximum likelihood
estimates.

\begin{figure}[h]
\begin{center}
\includegraphics[width=140mm,angle=0,clip=]{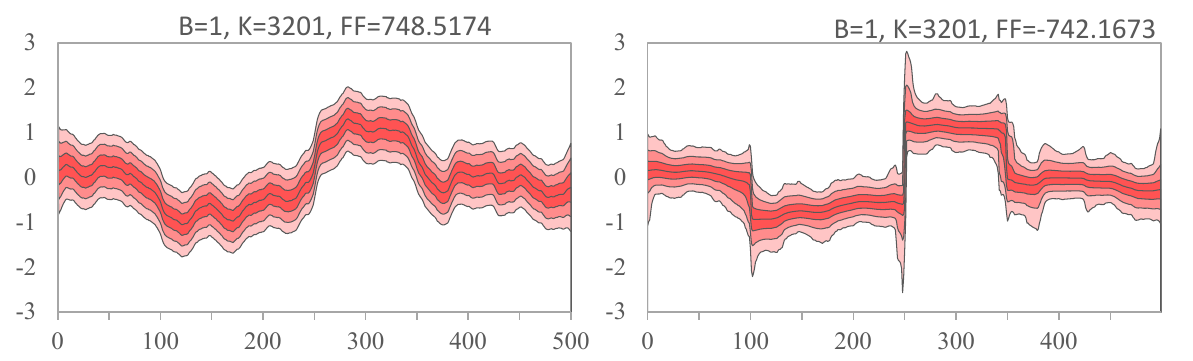}
\end{center}
\caption{Posterior summaries of the smoothed trend distributions obtained
from the one-dimensional trend model. Left: Gaussian system-noise model.
Right: Cauchy system-noise model.}
\label{Fig_MLE_Gauss_and_Cauchy}
\end{figure}

The left panel of Figure~\ref{Fig_MLE_Gauss_and_Cauchy} shows the
posterior mean together with the intervals corresponding to
$\pm1$, $\pm2$, and $\pm3$ posterior standard deviations at each time
point for the Gaussian model, corresponding to the limiting case
$b\rightarrow\infty$. The right panel shows the corresponding posterior
quantile intervals for the Cauchy model, for which $b=1$.

Under the Gaussian model, the estimated trend changes gradually over
time in response to the level shifts in the observations. In contrast,
under the Cauchy model, the estimated trend remains nearly constant
between the three change points and exhibits abrupt transitions near
those points. Thus, the heavy-tailed system-noise distribution produces
a piecewise-stable trend estimate that closely reflects the underlying
level shifts.

\begin{figure}[tbp]
\begin{center}
\includegraphics[width=84mm,angle=0,clip=]{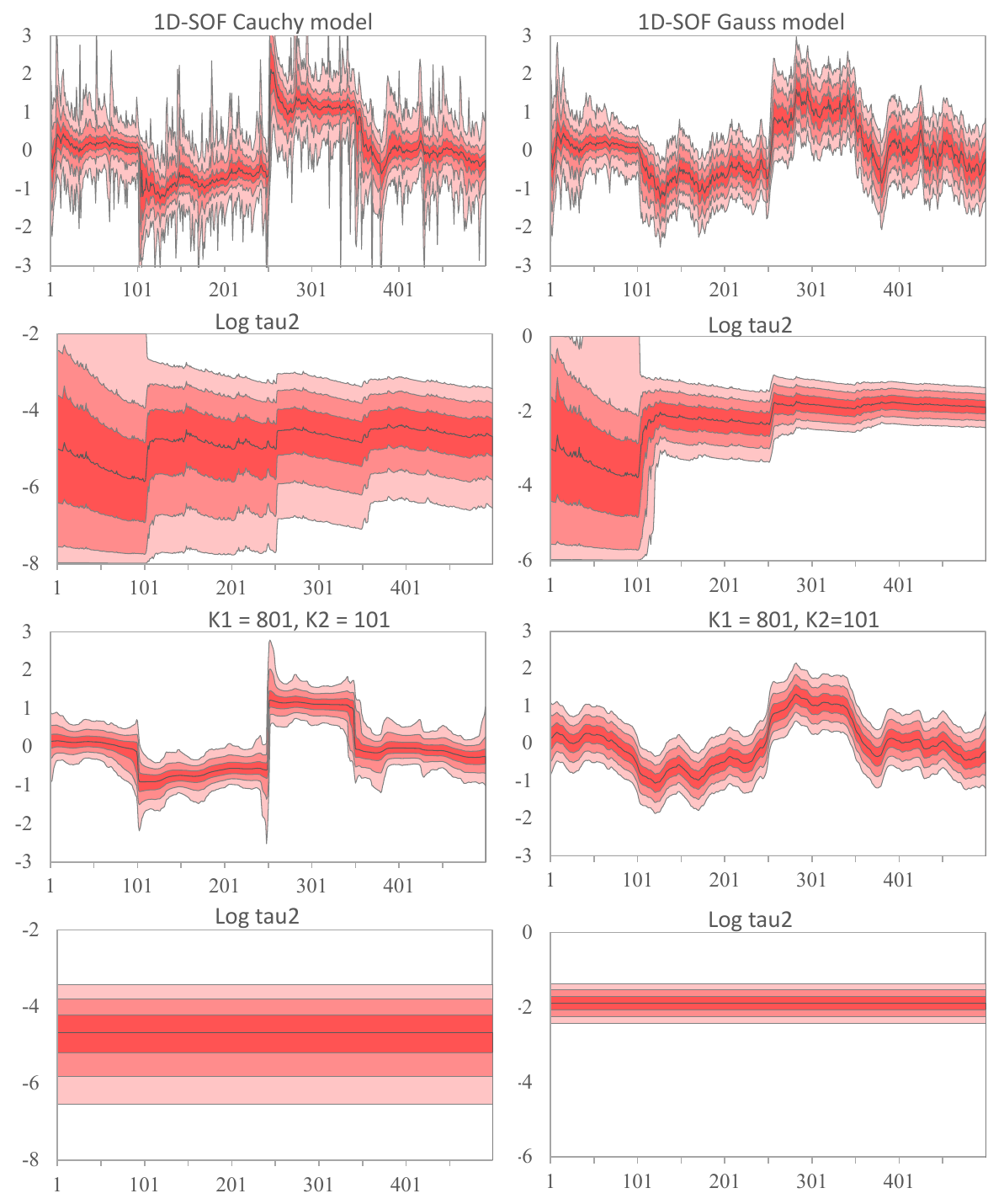}
\includegraphics[width=84mm,angle=0,clip=]{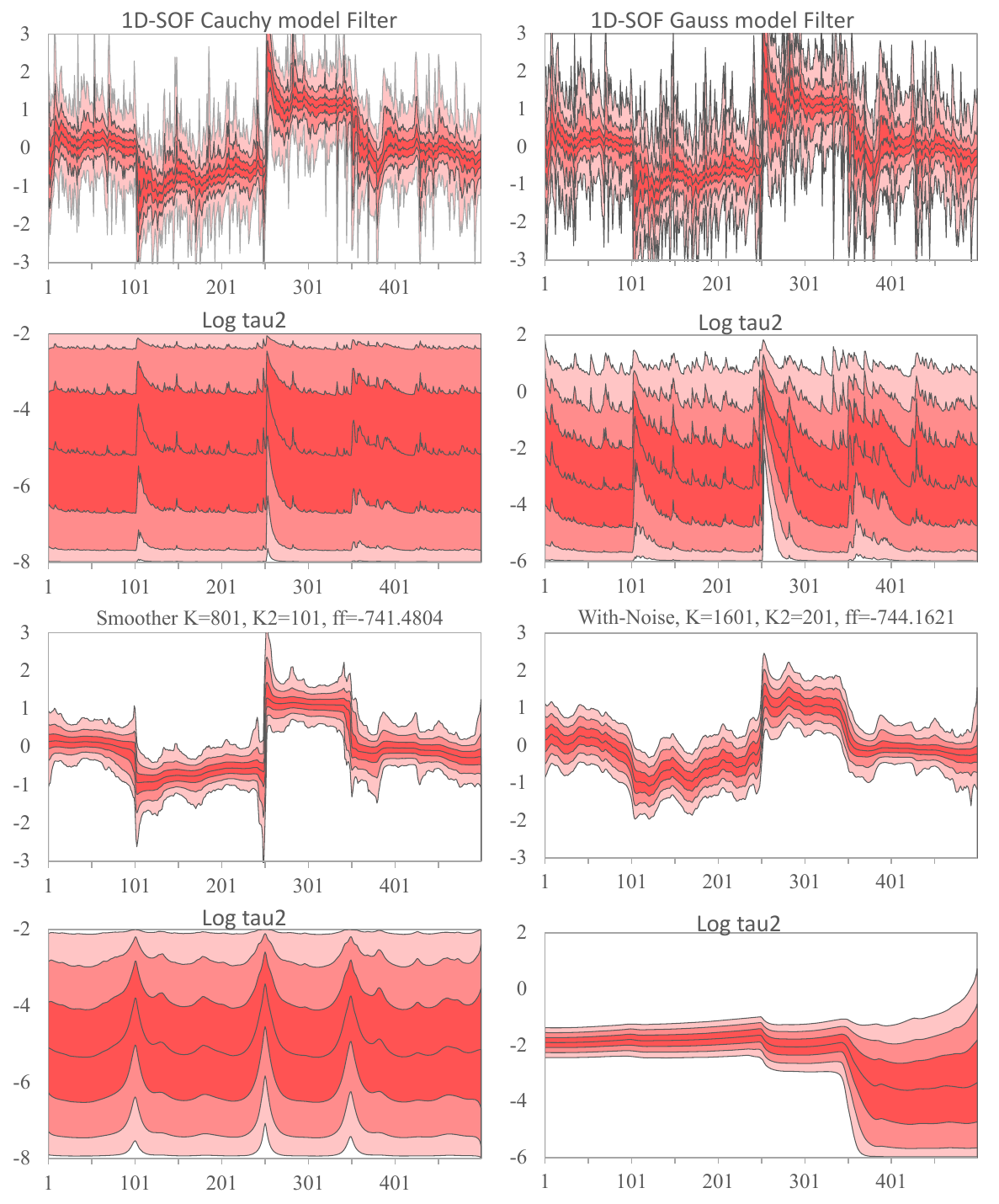}
\end{center}
\caption{Posterior distributions obtained using a one-parameter
self-organizing state-space model. From left to right, the columns show
the results for the time-invariant Cauchy model, the time-invariant
Gaussian model, the time-varying Cauchy model, and the time-varying
Gaussian model. From top to bottom, the rows show the filtering
distributions of the trend and the parameter, followed by their
corresponding smoothing distributions.}
\label{Fig:1D-SOF_posterior}
\end{figure}

Figure~\ref{Fig:1D-SOF_posterior} shows the results obtained using a
one-parameter self-organizing state-space model (SO-SSM), in which only
the system-noise variance $\tau^2$ is treated as unknown. For numerical
convenience and to ensure positivity, the parameter is represented as
\begin{align}
   \theta_n=\log\tau_n^2.
\end{align}

We first consider the time-invariant parameter models shown in the first
and second columns. 
For both the Cauchy and Gaussian models, the filtering distributions of
$\log\tau_n^2$ are widely dispersed at the beginning of the observation
period, particularly for $n<100$, and gradually become more concentrated
as additional observations are incorporated.
Since $\tau^2$ is assumed to be constant, no
evolution noise is introduced into the parameter equation. Consequently,
the smoothed posterior distribution of $\log\tau_n^2$ is identical at
all time points and is equal to the filtering posterior distribution at
the final time point, $n=N$.

Furthermore, the smoothed posterior distribution of the trend is almost
identical to that shown in Figure~\ref{Fig_MLE_Gauss_and_Cauchy}, which
was obtained by fixing the parameter at its maximum likelihood estimate.
This result indicates that simultaneous Bayesian estimation yields
state estimates that are consistent with those obtained using the
maximum likelihood estimate.

Unlike a standard state-augmentation particle filter, the Non-Gaussian
Filter can directly accommodate a constant unknown parameter without
introducing artificial evolution noise. This also reduces the
computational cost because the convolution integral associated with the
parameter evolution is unnecessary when its evolution variance is zero.

The third and fourth columns of
Figure~\ref{Fig:1D-SOF_posterior} show the results obtained when the
parameter is allowed to vary over time according to
\begin{align}
   \theta_n
   &= \theta_{n-1}+u_n,
   \qquad
   u_n\sim\mathcal{N}(0,\zeta^2),
   \qquad
   \theta_n=\log\tau_n^2.
\end{align}
The evolution variance $\zeta^2$ is estimated by maximum likelihood.
The resulting estimates are $\zeta^2=0.36$ for the Cauchy model and
$\zeta^2=0.15$ for the Gaussian model.

For the Cauchy model, the posterior mode and other quantiles of
$\tau_n^2$ increase around the time points at which the trend exhibits
abrupt jumps. Nevertheless, the posterior distribution of the trend is
similar to that obtained under the time-invariant parameter model. This
indicates that the heavy-tailed Cauchy system noise can already
accommodate abrupt changes without requiring substantial temporal
variation in its scale parameter.

In contrast, for the Gaussian model, the posterior distribution of the
trend differs substantially from that obtained under the time-invariant
parameter model and somewhat resembles the result obtained under the
Cauchy model. The time-varying variance increases near the change
points, allowing the Gaussian model to respond rapidly to abrupt level
shifts. Thus, by allowing $\tau_n^2$ to vary over time, even the Gaussian
model can adapt effectively to structural changes in the data.


\begin{figure}[h]
\begin{center}
\includegraphics[width=85mm,angle=0,clip=]{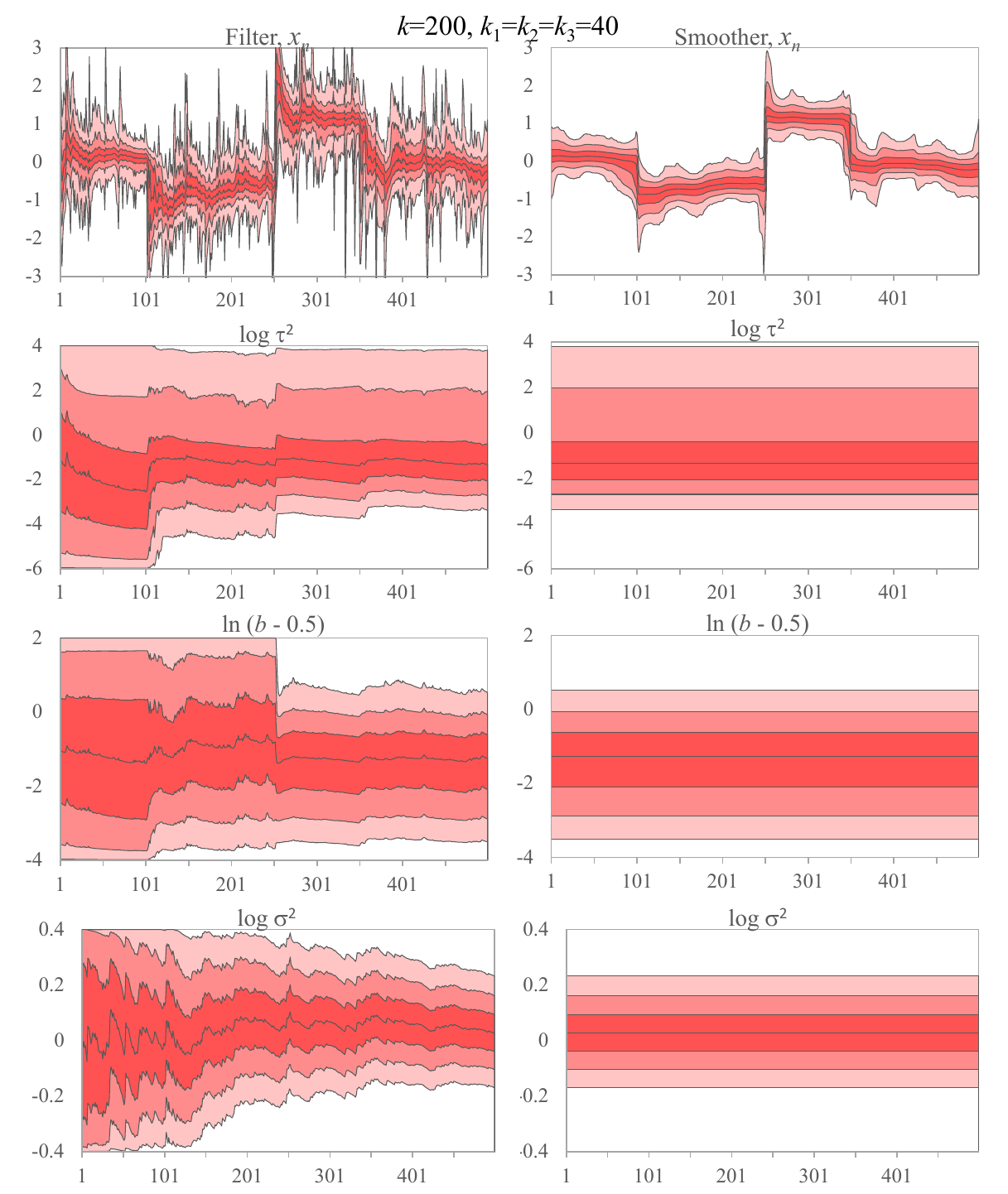}
\includegraphics[width=82mm,angle=0,clip=]{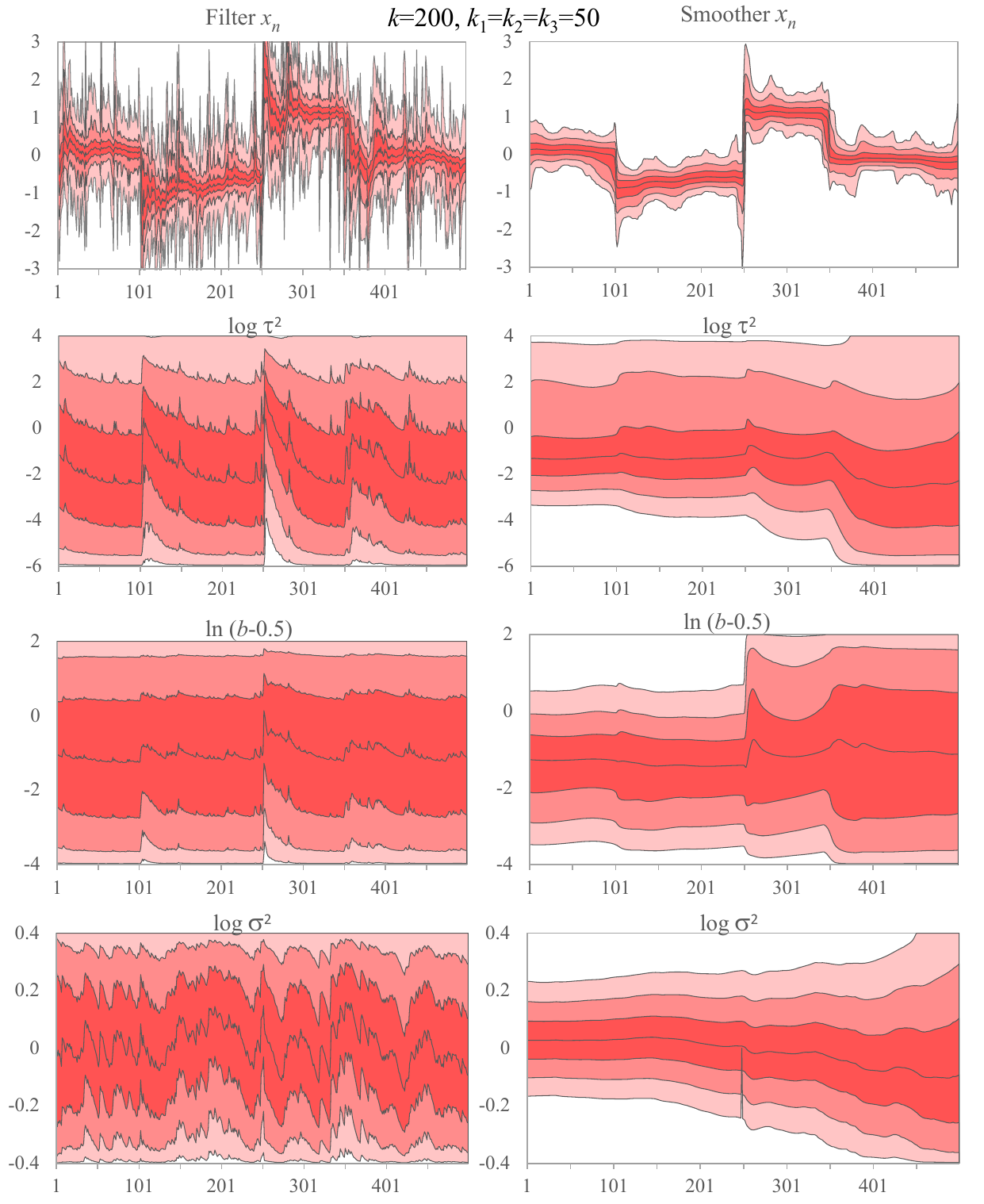}
\end{center}
\caption{Posterior distributions obtained using a three-parameter
self-organizing state-space model. The first and second columns show the filtering
and smoothing posterior distributions, respectively,
for the time-invariant parameter model.
The third and fourth columns show the corresponding
results for the time-varying parameter model. 
From top to bottom, the rows show the posterior distributions of
the trend, $\log_{10}\tau^2$, $\log(b-0.5)$, and $\log_{10}\sigma^2$.}
\label{Fig:3D-SOF_posterior}
\end{figure}

\subsection{Example: Three-Parameter Case}

Figure~\ref{Fig:3D-SOF_posterior} illustrates the simultaneous Bayesian
estimation of the latent state together with three unknown parameters
using the self-organizing state-space model. The unknown parameter vector is defined as
\begin{align}
  \theta_n
  =
  \left(
    \log_{10}\tau_n^2,\,
    \log(b_n-0.5),\,
    \log_{10}\sigma_n^2
  \right)^{\mathsf T}.
\end{align}
The logarithmic transformations ensure that
$\tau_n^2>0$, $b_n>0.5$, and $\sigma_n^2>0$.

Two models are considered. 
In the time-invariant model, the parameter vector is assumed to satisfy
\begin{align}
  \theta_n=\theta_{n-1}.
\end{align}
In the time-varying model, the parameters are allowed to evolve
according to a random-walk model:
\[
\theta_n=\theta_{n-1}+u_n,
\qquad
u_n\sim N(0,\zeta^2I).
\]

The first and second columns of
Figure~\ref{Fig:3D-SOF_posterior} show the filtering and smoothing
results, respectively, for the time-invariant parameter model.
As additional observations are incorporated, the filtering
distributions of the three parameters gradually become more
concentrated. Because $\tau^2$, $b$, and $\sigma^2$ are assumed to be
constant over time, their smoothed posterior distributions are identical
at all time points and are equal to the corresponding filtering
posterior distributions at the final time point, $n=N$.

The corresponding smoothed posterior distribution of the trend is obtained by
integrating over the posterior uncertainty in all three parameters.
These results demonstrate that the self-organizing state-space model
extends the deterministic non-Gaussian filter to a fully Bayesian framework.
The third and fourth columns of
Figure~\ref{Fig:3D-SOF_posterior} show the corresponding results for the
time-varying parameter model. For simplicity, the three parameter
components are assumed to follow independent Gaussian random walks with
a common evolution variance.

The posterior distributions of $\tau_n^2$ and $b_n$ indicate that their
estimated values tend to increase near the time points at which the
trend exhibits abrupt changes. The temporal behavior of the observation
noise variance $\sigma_n^2$ is more complex. The smoothed posterior
distributions of all three parameters vary gradually over time,
reflecting the assumed time-varying parameter dynamics.

The resulting smoothed posterior distribution of the trend is
similar to that obtained by fixing the parameters at their maximum
likelihood estimates, as shown in
Figure~\ref{Fig_MLE_Gauss_and_Cauchy}. Thus, allowing the parameters to
vary over time does not substantially alter the estimated trend in this
example, although the Bayesian approach provides additional information about temporal
changes in the system-noise and observation-noise distributions.

Maximum likelihood estimation provides only point estimates of the
unknown parameters. In contrast, the self-organizing state-space model
provides their joint posterior distribution together with the posterior
distribution of the latent state. 
The self-organizing state-space model therefore provides a fully Bayesian
characterization of both parameter uncertainty and state-estimation uncertainty.

These results demonstrate that the proposed
self-organizing state-space model naturally
extends deterministic likelihood-based estimation
to a fully Bayesian framework while preserving
essentially the same state estimates.

\section{Concluding Remarks}

This paper has demonstrated that the non-Gaussian filter provides a
practical deterministic framework for both maximum likelihood and
Bayesian inference in state-space models.

When the non-Gaussian filter was first proposed in 1987, its practical
use was severely restricted by computational cost and memory
requirements, except for simple models. 
The dramatic advances in processor speed and memory capacity over the
past four decades have fundamentally changed this situation. 
For state-space models of low- to moderate-dimension, 
the method can now be implemented on personal computers and 
applied to problems that were previously computationally prohibitive.

Whenever the non-Gaussian filter is computationally feasible, it has important
advantages for likelihood-based parameter estimation. The likelihood
computed by a particle filter is subject to Monte Carlo variability,
producing an irregular objective function and complicate
numerical optimization. In contrast, the non-Gaussian filter provides a
smooth and deterministic likelihood function. The resulting log-likelihood
can therefore be maximized reliably using standard numerical
optimization methods, even when a relatively modest number of grid
points is used.

The non-Gaussian filter also provides a natural framework for Bayesian
state and parameter estimation through the self-organizing state-space
model. In a standard state-augmented particle filter, fixed unknown
parameters are difficult to estimate because the associated particles
tend to degenerate. Artificial evolution noise is therefore often
introduced to maintain particle diversity, effectively treating the
parameters as time varying. The non-Gaussian filter can instead handle
fixed parameters directly, without introducing artificial parameter
dynamics, while naturally accommodating genuinely time-varying parameters
when required.

The principal limitation of the non-Gaussian filter remains the curse
of dimensionality. If each component of a $d$-dimensional continuous
state vector is represented by $k$ grid points, the number of grid
points increases exponentially with $d$, and the computational burden
rapidly becomes prohibitive. The method is therefore most suitable for
models with low- to moderate-dimensional continuous state vectors.
Nevertheless, its practical range of applicability can be extended 
when different numbers of grid points are used for different state
variables or when some components are discrete and take only a small
number of possible values.

The non-Gaussian filter should therefore no longer be regarded as a
computationally obsolete method. Advances in computing technology have
transformed it into a practical deterministic alternative to
simulation-based methods for a substantial class of state-space models.
For problems in which deterministic numerical integration remains
computationally manageable, it provides a particularly reliable framework for
likelihood evaluation, maximum likelihood estimation, and Bayesian
inference. 
The non-Gaussian filter and the particle filter should therefore be
viewed as complementary rather than competing methodologies. The former
is particularly attractive when accurate deterministic likelihood
evaluation is required, whereas the latter remains indispensable for
high-dimensional state estimation where grid-based numerical
integration is computationally infeasible.


\newpage
\appendix

\section{Kalman Filter and Exact Log-Likelihood}

For completeness, we briefly summarize the Kalman filter and the exact
log-likelihood for a linear Gaussian state-space model. These formulas
are used in Section~3.1 to evaluate the exact likelihood for the linear
Gaussian trend model.

Consider the linear Gaussian state-space model
\begin{align}
  x_n
  &= F_n x_{n-1}+G_n v_n,
  \qquad
  v_n\sim\mathcal{N}(0,Q_n),
  \label{Gaussian SSM-1}
  \\
  y_n
  &= H_n x_n+w_n,
  \qquad \hspace{7mm}
  w_n\sim\mathcal{N}(0,R_n),
  \label{Gaussian SSM-2}
\end{align}
where \(x_n\) denotes the latent state vector and \(y_n\) denotes the
observation. The initial state is assumed to follow
\(  x_0\sim\mathcal{N}(x_{0|0},V_{0|0}) \).
The system noise \(v_n\), the observation noise \(w_n\) and the initial
state \(x_0\) are assumed to be mutually independent.
Let \(  Y_{1:n}=\{y_1,\ldots,y_n\}\).
The one-step-ahead prediction mean and covariance matrix are denoted by
\(x_{n|n-1}\) and \(V_{n|n-1}\), respectively, and the filtering mean
and covariance matrix are denoted by \(x_{n|n}\) and \(V_{n|n}\).
They are computed recursively by the Kalman filter
(Anderson and Moore, 2012; Kitagawa, 2020).

\paragraph{Prediction step.}
\begin{align}
  x_{n|n-1}
  &= F_n x_{n-1|n-1},
  \nonumber\\
  V_{n|n-1}
  &= F_n V_{n-1|n-1}F_n^{\mathsf T}
     +G_nQ_nG_n^{\mathsf T}.
  \label{Eq-3-2}
\end{align}

\paragraph{Filtering step.}
Define the one-step-ahead prediction error and its variance by
\begin{align}
  \varepsilon_n
  &= y_n-H_nx_{n|n-1},
  \nonumber\\
  r_n
  &= H_nV_{n|n-1}H_n^{\mathsf T}+R_n.
  \label{Eq_prediction_error}
\end{align}
The Kalman gain and the filtering mean and covariance matrix are then
given by
\begin{align}
  K_n
  &=V_{n|n-1}H_n^{\mathsf T}r_n^{-1},
  \nonumber\\
  x_{n|n}
  &=x_{n|n-1}+K_n\varepsilon_n,
  \label{Eq-3-3}\\
  V_{n|n}
  &=(I-K_nH_n)V_{n|n-1}.
  \nonumber
\end{align}

For a parameter vector \(\theta\), the likelihood of the observed time
series \(Y_{1:N}\) can be decomposed into the product of one-step-ahead
predictive densities:
\begin{align}
  L(\theta)
  &=p(Y_{1:N}\mid\theta)
    =\prod_{n=1}^{N}
      p(y_n\mid Y_{1:n-1},\theta).
\end{align}
For the scalar-observation model considered here, the predictive
distribution is Gaussian:
\begin{align}
  g_n(y_n\mid Y_{1:n-1},\theta)
  &=
  p(y_n\mid Y_{1:n-1},\theta)
  \nonumber\\
  &=
  \frac{1}{\sqrt{2\pi r_n}}
  \exp\left(
    -\frac{\varepsilon_n^2}{2r_n}
  \right).
  \label{eq_distribution_g}
\end{align}
Consequently, the exact log-likelihood is
\begin{align}
  \ell(\theta)
  &=\log L(\theta)
  \nonumber\\
  &=
  -\frac{1}{2}
  \sum_{n=1}^{N}
  \left\{
    \log(2\pi)
    +\log r_n
    +\frac{\varepsilon_n^2}{r_n}
  \right\}
  \nonumber\\
  &=  -\frac{1}{2}  \left\{ N\log(2\pi)
    +\sum_{n=1}^{N}\log r_n
    +\sum_{n=1}^{N}\frac{\varepsilon_n^2}{r_n}
  \right\}.
  \label{Eq_log-lk}
\end{align}
The maximum likelihood estimate is obtained by maximizing
\(\ell(\theta)\) with respect to \(\theta\).


\section{Particle Filter and Approximate Log-Likelihood}

For comparison, we briefly summarize the bootstrap particle filter used
in the numerical examples. The particle filter represents the filtering
and predictive distributions by finite sets of particles
(Gordon et al., 1993; Kitagawa, 1996, 2014).

Let
\begin{align}
  \{f_n^{(1)},\ldots,f_n^{(m)}\}
  &\sim p(x_n\mid Y_{1:n}),  \nonumber\\
  \{p_n^{(1)},\ldots,p_n^{(m)}\}
  &\sim p(x_n\mid Y_{1:n-1})
\end{align}
denote the filtering and one-step-ahead predictive particles,
respectively. The corresponding distributions are approximated by
empirical distributions supported on these particles.

For the nonlinear non-Gaussian state-space model
\begin{align}
  x_n  &=f(x_{n-1})+g(v_n),  \qquad v_n\sim q(v),  \nonumber\\
  y_n  &=h(x_n)+w_n,
\end{align}
let \(r(y_n-h(x_n))\) denote the conditional observation density
\(p(y_n\mid x_n)\). The bootstrap particle filter proceeds as follows.

\begin{enumerate}
\item Generate the initial particles
\[
  f_0^{(j)}\sim p_0(x),
  \qquad j=1,\ldots,m,
\]
and assign equal normalized weights
\[
  \widetilde{w}_0^{(j)}=\frac{1}{m}.
\]

\item For \(n=1,\ldots,N\), repeat the following steps:
  \begin{enumerate}

  \item Generate independent system-noise samples
  \[
    v_n^{(j)}\sim q(v),
    \qquad j=1,\ldots,m.
  \]

  \item Generate the predictive particles
  \[
    p_n^{(j)}  =
    f\!\left(f_{n-1}^{(j)}\right) +g\!\left(v_n^{(j)}\right),
    \qquad j=1,\ldots,m.
  \]

  \item Compute the unnormalized observation weights
  \[
    w_n^{(j)}
    =
    r\!\left(y_n-h(p_n^{(j)})\right),
    \qquad j=1,\ldots,m.
  \]

  \item Normalize the weights:
  \[
    \widetilde{w}_n^{(j)}
    =
    \frac{w_n^{(j)}}
    {\displaystyle\sum_{i=1}^{m}w_n^{(i)}},
    \qquad j=1,\ldots,m.
  \]

  \item If resampling is required, generate
  \(f_n^{(1)},\ldots,f_n^{(m)}\) independently from the empirical
  distribution
  \[
    \sum_{i=1}^{m}
    \widetilde{w}_n^{(i)}
    \delta_{p_n^{(i)}}(x),
  \]
  where \(\delta_a(x)\) denotes the point mass at \(a\). After
  resampling, reset the weights to \(1/m\).

  If resampling is not performed, retain the weighted particles
  \(\{p_n^{(j)},\widetilde{w}_n^{(j)}\}_{j=1}^{m}\) as the filtering
  approximation.
  \end{enumerate}
\end{enumerate}

When resampling is performed at every time point, the one-step-ahead
predictive density is estimated by
\begin{align}
  \hat{p}(y_n\mid Y_{1:n-1},\theta)
  =  \frac{1}{m}  \sum_{j=1}^{m}  r\!\left(y_n-h(p_n^{(j)})\right).
\end{align}
The likelihood and log-likelihood are therefore estimated by
\begin{align}
  \hat{L}(\theta)
  &=
  \prod_{n=1}^{N}  \hat{p}(y_n\mid Y_{1:n-1},\theta),  \nonumber\\
  \hat{\ell}(\theta)
  &=
  \sum_{n=1}^{N}  \log
  \biggl\{
    \frac{1}{m} \sum_{j=1}^{m} r\!\left(y_n-h(p_n^{(j)})\right)
  \biggr\}.
  \label{Eq:PF-log-likelihood}
\end{align}
Unlike the exact likelihood obtained by the Kalman Filter or the
deterministic likelihood obtained by the non-Gaussian filter,
\(\hat{\ell}(\theta)\) depends on the random particles and therefore
contains Monte Carlo variability.

In practice, resampling may be performed only when the effective sample
size falls below a prescribed threshold. Using the normalized weights,
the effective sample size is defined as
\begin{align}
  \operatorname{ESS}_n
  =
  \biggl\{
    \sum_{j=1}^{m}
    \left(\widetilde{w}_n^{(j)}\right)^2
  \biggr\}^{-1}.
\end{align}
For example, resampling is performed when
\[
  \operatorname{ESS}_n<0.5m.
\]\end{document}